\documentclass[11pt,a4paper]{article}

\usepackage[margin=1in]{geometry}

\usepackage{amsthm}
\usepackage{amsmath}

\usepackage{newtxtext}
\usepackage{newtxmath}

\usepackage{booktabs}
\usepackage{longtable}
\usepackage{tabularx}
\usepackage{array}
\usepackage{calc}
\newcommand{\tableinterval}[3][9]{%
  \ensuremath{[\makebox[\widthof{$-#1.999$}][r]{$#2$},\,
                 \makebox[\widthof{$-#1.999$}][r]{$#3$}]}%
}
\newcolumntype{Y}{>{\raggedright\arraybackslash}X}

\usepackage{comment}

\usepackage{setspace}
\usepackage{natbib}
\setcitestyle{authoryear,round}

\usepackage[english]{babel}
\usepackage{graphicx}
\usepackage{float}
\usepackage[section]{placeins}
\usepackage{chngcntr}
\usepackage{caption}
\usepackage{subcaption}

\usepackage{tcolorbox}
\usepackage{tikz}
\usetikzlibrary{matrix}

\usepackage[shortlabels]{enumitem}
\usepackage{etoolbox}
\usepackage{hyperref}

\AtBeginDocument{%
  \setlength{\abovedisplayskip}{9pt plus 2pt minus 2pt}%
  \setlength{\belowdisplayskip}{9pt plus 2pt minus 2pt}%
  \setlength{\abovedisplayshortskip}{5pt plus 2pt}%
  \setlength{\belowdisplayshortskip}{5pt plus 2pt minus 1pt}%
}

\AtBeginEnvironment{thebibliography}{\singlespacing\setlength{\bibsep}{5pt plus 1pt minus 1pt}}

\makeatletter

\newcommand{\Rmnum}[1]{\expandafter\@slowromancap\romannumeral #1@}
\makeatother

\theoremstyle{definition}

\newtheorem{proposition}{Proposition}

\title{When Does Learning Improve Correlation Mixtures for Value-at-Risk and Expected Shortfall?}
\author{
	Yifan He\thanks{
	Department of Mathematics and Statistics, Texas Tech University, Lubbock, Texas 79409, USA\\
	Email: \href{mailto:Yifan.He@ttu.edu}{Yifan.He@ttu.edu}}
 \and
 Svetlozar T. Rachev\thanks{Department of Mathematics and Statistics,
 Texas Tech University, Lubbock, Texas 79409, USA\\
Email: \href{mailto:Zari.Rachev@ttu.edu}{Zari.Rachev@ttu.edu}}}

\begin{document}
\date{}
\maketitle
\begin{abstract}
\noindent Does learning the weights of two correlation estimators improve one-day Value-at-Risk and Expected Shortfall forecasts over equal mixing? Using a retrospective evaluation of four portfolios of exchange-traded funds over 3,772 dates, we find that the incremental improvement from learning mixture weights over equal mixing is statistically inconclusive, even for the stock--bond portfolio, where the learned mixture substantially improves forecast accuracy relative to the equally weighted correlation estimator. These findings distinguish the benefits of combining correlation estimators from the additional value of learning their weights. We establish that two conditional loss laws can have identical endpoint forecasts and expected endpoint-score differences yet different optimal interior mixture weights. We further characterize the population-optimal weight along the variance-mixture forecast ray and examine direct-score training under common margins. In a known-margin simulation with fixed unconditional state probabilities, stronger state information improves the relative realized forecast performance of a ridge learner with driver interactions, whereas the corresponding gains for a boosted-stump learner are inconclusive under normal innovations and deteriorate under Student-$t_5$ innovations. Even with exact knowledge of the conditional endpoint-score difference, the sigmoid weighting rule incurs substantial expected regret relative to the feasible optimum. These results highlight the importance of distinguishing signal quality, weight selection, and feasible-family restrictions when evaluating adaptive forecast combinations.\par\medskip\noindent
\par\smallskip
\end{abstract}

\section{Introduction}\label{sec:introduction}
Exponential weighting can help a correlation estimator respond to changes in diversification, at the cost of greater sensitivity to recent observations. A mixture of equally and exponentially weighted estimators can moderate this trade-off. The question is whether a weight learned from historical market conditions improves one-day portfolio Value-at-Risk (VaR) and Expected Shortfall (ES) forecasts beyond a simple mixture.

We address this question by holding the marginal models and portfolio weights fixed across the dependence specifications. Both correlation estimators use standardized residuals from the same generalized autoregressive conditional heteroskedasticity (GARCH) models with Student-$t$ innovations. A shallow boosted-tree model predicts the difference in the joint Fissler--Ziegel score (FZ0) between the two estimators and converts that prediction to a correlation-mixture weight. Equal mixing provides a benchmark for the benefit of combining estimators. A fixed weight selected in validation and a rule based on the expanding mean score difference provide benchmarks for the additional value of market-state features.

We evaluate these forecasts retrospectively on four portfolios of exchange-traded funds (ETFs). For the stock--bond portfolio, the learned mixture improves joint VaR--ES accuracy relative to the equally weighted correlation estimator. The estimated gain over a mixture that gives both estimators equal weight is much smaller. Across portfolios and estimation windows, the evidence for an incremental gain from learning the weight is inconclusive.

A second question is whether the expected score gap between the two endpoint forecasts determines the optimal mixture weight. The joint score depends nonlinearly on the weight because the mixture acts on portfolio variance. We construct two conditional loss laws with the same endpoint forecasts and expected score gap but different optimal interior weights. When the endpoint variances differ, we characterize the unique population-optimal weight, including when it lies at an endpoint.

We also compare the learned mixture with a dynamic conditional correlation (DCC) benchmark and evaluate matched rules trained on either the mixture's FZ0 score or the endpoint-score gap. A known-margin simulation varies the predictability of the correlation state while holding its unconditional frequency fixed. It separates the score effects of estimating the gap, converting it to a weight, and restricting forecasts to the feasible mixture family.

The rest of the paper is organized as follows. Section~\ref{sec:literature-review} reviews related work, and Section~\ref{sec:data-and-experimental-design} describes the data and experimental design. Section~\ref{sec:models} presents the forecasting models and theoretical results, and Section~\ref{sec:evaluation} sets out the forecast evaluation methods. Section~\ref{sec:empirical} reports the empirical results, and Section~\ref{sec:economics} summarizes the exploratory buffer-cost comparison. Section~\ref{sec:simulation} presents the simulations. Section~\ref{sec:robustness} discusses computational cost and limitations, and Section~\ref{sec:conclusion} concludes.

\section{Literature Review}\label{sec:literature-review}
\subsection{Joint Tail-Risk Forecasts and Calibration}
Expected Shortfall measures the severity of losses in the tail and satisfies the coherence properties studied by \citet{artzner1999coherent} and \citet{acerbi2002coherence}. The joint elicitability of VaR and ES permits their comparison using consistent scoring functions \citep{fissler2016higher,patton2019dynamic}. \citet{taylor2019forecasting} links consistent joint evaluation to a semiparametric forecasting approach based on the asymmetric Laplace distribution.

Comparative accuracy and calibration address different properties of a forecast \citep{nolde2017elicitability}. Coverage and dynamic-quantile tests assess exceedance behavior \citep{kupiec1995techniques,christoffersen1998evaluating,engle2004caviar}, while joint identification moments and ES regressions assess restrictions involving tail severity \citep{bayer2022regression}. Conditional predictive ability provides a related perspective on how relative forecasting performance may vary with information available at the forecast origin \citep{giacomini2006tests}.

\subsection{Dependence Responsiveness and Forecast Combinations}
GARCH models describe conditional variance dynamics \citep{bollerslev1986generalized}. Exponential weighting \citep{riskmetrics1996} and dynamic conditional correlation \citep{engle2002dynamic} offer different ways of allowing dependence estimates to respond to new data. We study a fixed-decay exponentially weighted estimator alongside an equally weighted rolling estimator. Both use the same marginal forecasts, whose estimation uncertainty is relevant to the resulting risk measures \citep{christoffersen2005estimation}.

Evidence on the benefits of multivariate models varies across applications. \citet{fortin2023forecasting} find that univariate methods often match multivariate factor-based ES forecasts for equity portfolios. Other applications document gains from regime-dependent and semiparametric specifications \citep{fulle2024predicting,storti2025semiparametric}. This variation motivates a comparison that holds the marginal forecasts fixed and varies the way residual correlations are estimated and combined.

Forecast combination dates to \citet{bates1969combination}. The reviews by \citet{clemen1989combining} and \citet{wang2023forecastreview} trace the development from simple averages to estimated and time-varying weights. Equal weighting is a substantive benchmark because estimation error can offset the benefits of choosing weights from data. \citet{claeskens2016puzzle} analyze the bias and variance induced by estimated weights, \citet{blanc2016simple} examine the roles of sample size and structural change, and \citet{elliott2026averaging} characterize settings in which the gains from optimal linear weights are small.

For tail risk, \citet{bayer2018combining} combines VaR forecasts with penalized quantile regression, and \citet{taylor2020combinations} estimates VaR--ES combination weights using joint scores. \citet{wang2025tailrisk} use machine learning to estimate tail densities and study regularized VaR--ES combinations. \citet{dimitriadis2021encompassing} use joint scores to test whether one ES forecast encompasses information from another, and \citet{taylor2026combining} examine regularized combinations drawn from large pools of VaR--ES forecasts.

\subsection{Learning Weights for Correlation Mixtures}
Machine-learning approaches to tail risk include neural architectures \citep{qiu2024stateful} and generalized quantile forests that use economic and financial predictors \citep{james2026forest}. \citet{barrera2026statistical} establish non-asymptotic convergence results for two-step conditional VaR and ES estimation, including under heavy-tailed losses. Closest to our weighting question, \citet{tetereva2026combination} use trees to construct direct VaR--ES combinations with weights that vary across macro-financial regimes.

We use boosted trees \citep{chen2016xgboost} to combine residual-correlation estimators before projecting to portfolio VaR and ES. This placement changes the weight-selection problem: a predictor may discriminate between the two endpoint forecasts without identifying the interior correlation mixture that minimizes the joint score. Our analysis studies this distinction under common margins and uses equal-weight and feature-free controls to assess the additional value of state-dependent weighting.

\section{Data and Experimental Design}\label{sec:data-and-experimental-design}
\subsection{Data and Portfolio Construction}\label{sec:data_sources}
We use daily adjusted closing prices from Yahoo Finance for eight US-listed ETFs: SPY (US large-cap equities), IWM (US small-cap equities), EFA (developed-market foreign equities), EEM (emerging-market equities), IEF (intermediate Treasury bonds), TLT (long Treasury bonds), LQD (investment-grade corporate bonds), and IYR (US real estate).

Simple returns from adjusted prices proxy for gross total returns:
\begin{equation*}
 r_{i,t}=\frac{P^{\rm adj}_{i,t}}{P^{\rm adj}_{i,t-1}}-1.
\end{equation*}
The common sample contains 5,716 daily returns from April 15, 2003 through December 31, 2025. All prices are in US dollars. The foreign-equity ETFs therefore reflect currency exposure as well as fund-specific tracking effects.

We use the observed prices without interpolation, forward filling or winsorization.

Table~\ref{tab:weights} gives the target weights for four portfolios. For portfolio $p$, let $w_p$ contain its nonzero target weights, $\mathbf r_t$ the corresponding constituent returns, and $d$ the number of constituents. Portfolios are rebalanced daily to the target weights. The gross return is $r_{p,t}=w_p^{\prime}\mathbf r_t$, and the loss is $L_{p,t}=-r_{p,t}$. P1 is a stock--bond reference portfolio. P2 adds foreign equities and intermediate Treasuries, P3 adds corporate bonds, and P4 includes small-cap equities and real estate. Portfolio composition and dimension vary together.

\begin{table}[!htbp]
\centering\caption{Daily target portfolio weights.}\label{tab:weights}
\begingroup\small\singlespacing
\begin{tabular}{lrrrrrrrrr}\toprule
Portfolio & SPY & IWM & EFA & EEM & IEF & TLT & LQD & IYR & $d$\\\midrule
P1 & 0.60 & 0.00 & 0.00 & 0.00 & 0.00 & 0.40 & 0.00 & 0.00 & 2\\
P2 & 0.30 & 0.00 & 0.15 & 0.15 & 0.20 & 0.20 & 0.00 & 0.00 & 5\\
P3 & 0.30 & 0.00 & 0.15 & 0.15 & 0.15 & 0.15 & 0.10 & 0.00 & 6\\
P4 & 0.20 & 0.10 & 0.15 & 0.15 & 0.10 & 0.10 & 0.10 & 0.10 & 8\\\bottomrule
\end{tabular}\endgroup
\end{table}

\subsection{Estimation, Validation and Evaluation}\label{sec:chronology}
The pre-2008 return history provides initial inputs for the risk estimators. Each endpoint forecast uses the latest $W$ returns through $t-1$, with $W=756$ in the main analysis and $504$ and $1,008$ in sensitivity analyses. Marginal and pooled-tail parameters are re-estimated on the first available forecast date of each calendar month. Between refits, the current trailing window is filtered using the most recent successful estimates.

The learner uses 505 feature--target records from 2008--2009 for initial training and 252 dates in 2010 for prequential validation. Each monthly fit uses earlier observations, and the mean validation score determines the specification. During evaluation (3,772 dates from January 3, 2011 through December 31, 2025), the specification stays fixed and the model is refitted monthly on an expanding sample. The first evaluation fit uses 757 records. Earlier endpoint scores initialize the lagged-score features.

All adaptive comparisons use the 97.5\% confidence level and common evaluation dates. The window sensitivity analysis compares the two correlation estimators, the learned mixture, equal mixing, the expanding-mean gate and the validation-selected fixed mixture. The eleven-model confidence set uses the 756-day window.

Margin and common-tail re-estimation, M7 candidate selection, gap normalization, and weight fitting are performed by portfolio and estimation window. DCC and the matched state rules are evaluated at $W=756$ with fixed specifications. Appendix~\ref{app:implementation} records the predictor and label cutoffs and the results of truncated-data replays. The empirical evaluation is retrospective because the historical outcomes had been examined during research development. The P1 comparison with the equally weighted estimator and its ten-day bootstrap specification were set before generating those forecasts. The portfolio--window and mechanism comparisons are exploratory.

\section{Risk Forecasting Models}\label{sec:models}
For a given portfolio, we suppress the portfolio index $p$ and write $L_t=L_{p,t}$. Let $\mathcal F_{t-1}$ be the information available through date $t-1$. For integrable portfolio loss and confidence level $c\in(0,1)$, one-day-ahead conditional VaR and ES are defined by
\begin{align*}
 \operatorname{VaR}_c(L_t\mid\mathcal F_{t-1})
 &:=\inf\bigl\{x\in\mathbb R:
       \Pr(L_t\leq x\mid\mathcal F_{t-1})\geq c\bigr\},\\
 \operatorname{ES}_c(L_t\mid\mathcal F_{t-1})
 &:=\frac{1}{1-c}\int_c^1
       \operatorname{VaR}_u(L_t\mid\mathcal F_{t-1})\,du.
\end{align*}
Thus VaR is the conditional loss quantile and ES is the average of the upper-tail quantiles. For fixed $c$ and $\alpha=1-c$, abbreviate these true conditional risks as
\begin{equation*}
 q_t^{\ast}:=\operatorname{VaR}_c(L_t\mid\mathcal F_{t-1}),\qquad
 e_t^{\ast}:=\operatorname{ES}_c(L_t\mid\mathcal F_{t-1}).
\end{equation*}
For continuous conditional loss distributions, ES equals the conditional mean loss beyond VaR. Write $q_{j,t},e_{j,t}$ for model $j$'s one-day-ahead forecasts and $q_t(g),e_t(g)$ for forecasts at mixture weight $g$. We suppress the portfolio index on risks, variances and scores. The parametric models use symmetric innovations and a zero conditional return mean. Historical simulation retains the location of the empirical return sample.

\subsection{Univariate Benchmark Models}\label{subsec:univariate-benchmarks}
\paragraph*{Historical Simulation (M1).}
M1 uses historical simulation (HS). It forecasts VaR, $q_{1,t}$, by inverting the empirical cumulative distribution function (CDF) of losses. The ES forecast integrates that distribution's upper-tail quantiles. If $\ell_{(1)}\geq\cdots\geq\ell_{(W)}$ are the ordered losses, $k=\lfloor W\alpha\rfloor$, and $\delta=W\alpha-k$, then
\begin{equation*}
 e_{1,t}=\frac{\sum_{j=1}^k\ell_{(j)}+\delta\ell_{(k+1)}}{W\alpha}.
\end{equation*}
The fractional contribution of the boundary observation gives the exact empirical tail average. The rolling window gives each included observation the same weight until it leaves the sample.

\paragraph*{Exponentially Weighted Moving Average (M2).}
For the univariate models, the model index on conditional variances and volatilities is suppressed when the model under discussion is clear. M2 uses Gaussian innovations and an exponentially weighted moving average (EWMA) variance estimate under a zero conditional mean. Within each trailing window, the first 75 squared returns initialize the second moment. Remaining observations update it as
\begin{equation*}
 h_{s+1}=\lambda h_s+(1-\lambda)r_{p,s}^2,\qquad \lambda=0.94.
\end{equation*}
We use the daily decay convention of RiskMetrics \citep{riskmetrics1996}. Here $\Phi$ and $\phi$ are the standard normal CDF and probability density function (PDF). With $\sigma_t=\sqrt{h_t}$ and $z_c=\Phi^{-1}(c)$,
\begin{equation*}
 q_{2,t}=\sigma_t z_c,\qquad
 e_{2,t}=\sigma_t\frac{\phi(z_c)}{\alpha}.
\end{equation*}

\paragraph*{GARCH-Based Models (M3 and M4).}
M3 and M4 estimate a zero-mean GARCH(1,1) on portfolio returns, where $z_t$ is the standardized innovation with zero mean and unit variance:
\begin{equation*}
 r_{p,t}=\sigma_t z_t,\qquad
 \sigma_t^2=\omega+a r_{p,t-1}^2+b\sigma_{t-1}^2,
\end{equation*}
with $\omega>0$, $a,b\geq0$, and $a+b\leq1$. M3 uses normal innovations and gives
\begin{equation*}
 q_{3,t}=\sigma_t z_c,\qquad
 e_{3,t}=\sigma_t\frac{\phi(z_c)}{\alpha}.
\end{equation*}
M4 uses variance-standardized Student-$t$ innovations with estimated $\nu>2$. Let $T_\nu$ and $f_\nu$ denote the CDF and PDF of the conventional Student-$t$ distribution with $\nu$ degrees of freedom. For $z_{c,\nu}=T_\nu^{-1}(c)$ and $s_\nu=\sqrt{(\nu-2)/\nu}$, its risk forecasts are
\begin{align*}
 q_{4,t}&=\sigma_t s_\nu z_{c,\nu},\\
 e_{4,t}&=\sigma_t s_\nu
 \frac{\nu+z_{c,\nu}^2}{\nu-1}\frac{f_\nu(z_{c,\nu})}{\alpha}.
\end{align*}
We estimate separate parameter sets for the normal and Student-$t$ specifications.

\subsection{Multivariate Risk Forecasting Models}\label{subsec:multivariate-risk-forecasting-models}
M5 and M6 fit a GARCH variance process with Student-$t$ innovations separately to each constituent asset. The asset-specific degrees of freedom enter the marginal variance fits. On refit dates, we estimate a common predictive degrees-of-freedom parameter $\nu\in[2.1,100]$ by pooling the marginal Student-$t$ log likelihoods of the variance-standardized residuals. We specify an elliptical multivariate Student-$t$ forecast with this common parameter and covariance $D_t R_t D_t$.

Let $z_s$ be the vector of filtered standardized residuals and define the correlation-normalization operator by
\begin{equation*}
 \operatorname{CorrNorm}(A)_{ij}:=\frac{A_{ij}}{\sqrt{A_{ii}A_{jj}}}.
\end{equation*}
M5 uses
\begin{equation}\label{eq:eqcorr}
 \widehat R^{\rm EQ}_t=\operatorname{CorrNorm}\left(W^{-1}\sum_{s=t-W}^{t-1}z_sz_s^{\prime}\right).
\end{equation}
The equally weighted (EQ) correlation matrix is updated daily. We use uncentered residual second moments, consistent with the zero-mean specification.

M6 instead uses the exponentially weighted (EW) residual second-moment matrix $A_s$, updated as
\begin{equation}\label{eq:ewcorr}
 A_{s+1}=0.94A_s+0.06z_sz_s^{\prime},\qquad
 \widehat R^{\rm EW}_t=\operatorname{CorrNorm}(A_t).
\end{equation}
For each forecast date $t$, initialize the recursion within the current trailing window by
\begin{equation*}
 A_{t-W+75}:=\frac{1}{75}
 \sum_{s=t-W}^{t-W+74}z_sz_s^{\prime}.
\end{equation*}
The initialization uses the uncentered second moment of the first 75 standardized residual vectors. The recursion for $s=t-W+75,\ldots,t-1$ gives $A_t$. The estimators in Equations~\eqref{eq:eqcorr} and \eqref{eq:ewcorr} share the same marginal forecasts, predictive $\nu$, estimation histories and portfolio weights. For $D_t=\operatorname{diag}(\widehat\sigma_{1,t},\ldots,\widehat\sigma_{d,t})$, let $\widehat R_t$ be the M5 or M6 correlation estimate. The corresponding portfolio variance is
\begin{equation}\label{eq:projection}
 \widehat\sigma_{p,t}^2=w_p^{\prime}D_t\widehat R_tD_tw_p.
\end{equation}
We obtain portfolio VaR and ES by applying the Student-$t$ tail formulas to the conditional standard deviation from Equation~\eqref{eq:projection}.

If a marginal GARCH fit fails after a retry, the most recent successful parameter estimates are retained. If any constituent asset lacks an initial successful marginal fit, both endpoint specifications temporarily issue the portfolio-level Gaussian EWMA forecasts from M2. The fallback issues VaR and ES directly from the portfolio return series. Once successful marginal fits are available for all constituent assets, the GARCH-standardized residuals are used to estimate dependence and pooled degrees of freedom for the common Student-$t$ projection. If the pooled degrees-of-freedom estimation fails, the most recent successful estimate is retained, or a value of eight is used for initialization. M5 and M6 each estimate $4d+1$ parameters.

\paragraph*{DCC Benchmark.}
We fit \citet{engle2002dynamic}'s scalar DCC recursion $Q_{s+1}=(1-a-b)\overline Q+a z_s z_s^{\prime}+bQ_s$, with $a,b\geq0$ and $a+b\leq0.999$. The current window's uncentered residual second moment $\overline Q$ initializes and targets the recursion. Gaussian correlation quasi-likelihood selects the coefficients at the first trading day of each month. Held fixed within the month, the coefficients are applied to the daily trailing residual window, yielding $\widehat R_t^{\rm DCC}=\operatorname{CorrNorm}(Q_t)$. The forecasts share the same standardized residuals, marginal volatilities, common Student-$t$ parameter, forecast dates and portfolio weights as M5 and M6.

\subsection{State-Dependent Correlation Mixing}\label{sec:gate}
Let $g_t\in[0,1]$ be the weight on M6 and retain $w_p$ for portfolio allocation. Set $R_{5,t}:=\widehat R^{\rm EQ}_t$ and $R_{6,t}:=\widehat R^{\rm EW}_t$, and define
\begin{equation*}
 v_{j,t}:=w_p^{\prime}D_tR_{j,t}D_tw_p,\qquad j\in\{5,6\}.
\end{equation*}
Thus $R_{j,t}$ is a forecast correlation matrix, $v_{j,t}$ is the corresponding portfolio conditional variance forecast, and $\sqrt{v_{j,t}}$ is its conditional standard deviation forecast.

Write $\nu_t$ for the common predictive degrees of freedom in force at date $t$. With $z_{c,\nu_t}=T_{\nu_t}^{-1}(c)$ and $s_{\nu_t}=\sqrt{(\nu_t-2)/\nu_t}$, the shared standardized Student-$t$ tail constants follow from the GARCH formulas in Section~\ref{subsec:univariate-benchmarks}:
\begin{align*}
 c_{q,t}&=s_{\nu_t}z_{c,\nu_t},\\
 c_{e,t}&=s_{\nu_t}\frac{\nu_t+z_{c,\nu_t}^{2}}{\nu_t-1}
                \frac{f_{\nu_t}(z_{c,\nu_t})}{\alpha}.
\end{align*}
The endpoint forecasts satisfy $(q_{j,t},e_{j,t})=\sqrt{v_{j,t}}(c_{q,t},c_{e,t})$ for $j\in\{5,6\}$. M7 uses
\begin{align}
 R_{7,t}&=(1-g_t)R_{5,t}+g_tR_{6,t},\notag\\
 v_{7,t}&=(1-g_t)v_{5,t}+g_tv_{6,t},\qquad
 (q_{7,t},e_{7,t})=\sqrt{v_{7,t}}(c_{q,t},c_{e,t}),\label{eq:mixtail}
\end{align}
Convex mixing preserves the unit diagonal and positive semidefiniteness. Both endpoints are checked for positive definiteness, which then holds throughout the mixture interval. By Equation~\eqref{eq:mixtail}, VaR and ES are nonlinear in the mixture weight.

Define the realized endpoint-score gap
\begin{equation*}
 \Delta S_t=S_t(q_{6,t},e_{6,t})-S_t(q_{5,t},e_{5,t}),
\end{equation*}
where $S_t$ is the joint FZ0 score, defined under the positive-loss convention in Equation~\eqref{eq:fz}. At monthly refit $m$, the extreme gradient boosting (XGBoost) regressor $f_m$ is trained on $\Delta S_s/s_m$ for records $s<m$, where $s_m$ is the sample standard deviation of the training targets. Let $X_{t-1}$ be the twelve features available through date $t-1$, and define $\operatorname{logistic}(x)=(1+\exp(-x))^{-1}$. The mixture weight is
\begin{equation*}
 g_t=\operatorname{logistic}[-f_m(X_{t-1})],\qquad t\text{ in month }m.
\end{equation*}
A negative predicted gap increases the weight on M6. The continuous target measures the magnitude as well as the direction of the endpoint-score difference.

The twelve features are the 21- and 63-day portfolio return standard deviations and their ratio, mean off-diagonal correlations for M5 and M6, root-mean-square disagreement between M5 and M6, disagreement between 21- and 252-day residual correlations, the change between adjacent 21-day residual correlations, the M6-to-M5 portfolio volatility ratio, and the 21- and 63-day means and 63-day sample standard deviation of the realized score gap. All feature windows end at $t-1$.

The four boosted-tree candidates combine maximum depths of one or two with 40 or 80 trees. They use a learning rate of 0.03, squared-error loss, histogram-based tree fitting, minimum child weight of 30 and an L2 leaf penalty of 20. We choose the candidate with the lowest mean prequential FZ0 in 2010 separately for each estimation window. Table~\ref{tab:adaptive_candidates} reports the main-window selections.

M7 adds the parameters of the weight function.

\subsection{Simple Controls and Mechanism Diagnostics}\label{sec:controls}
M8 fixes $g_t=0.5$. M9 chooses the EW endpoint when lagged 21-day volatility exceeds its monthly expanding-training 75th percentile. M10 chooses EW when the lagged 63-day score gap is negative. M11 has no market-state inputs:
\begin{equation*}
 g_{11,t}=\operatorname{logistic}(-\overline{\Delta S}_{m}/s_m),
\end{equation*}
where $\overline{\Delta S}_m$ is the mean realized endpoint-score gap over the expanding training sample available before refit $m$, and $s_m$ uses the same sample. M12 selects the weight that minimizes mean 2010 mixture FZ0 over $g\in\{0,0.05,\ldots,1\}$ and holds it fixed during evaluation. M8, M11 and M12 benchmark equal mixing, feature-free updating and validation-based weight estimation, respectively.

Two exploratory diagnostics use the same endpoints. M13 chooses an amplification coefficient $k\in\{0,0.25,0.5,1,2,4\}$ on the 2010 validation sample, replacing the predicted normalized gap by
\begin{equation*}
 \overline{\Delta S}_{m}/s_m+k\{f_m(X_{t-1})-\overline{\Delta S}_{m}/s_m\}.
\end{equation*}
At $k=0$, the weight depends only on the historical mean gap. At $k=1$, the rule recovers M7. In the second diagnostic, M14 directly minimizes historical mixture FZ0 in a linear sigmoid family, and M16 fits the normalized endpoint-score gap in the same family. M15 estimates a feature-free constant weight at each monthly refit. Appendix~\ref{app:implementation} gives the estimation details.

We also construct a matched four-state comparison. States cross whether lagged 21-day volatility exceeds its expanding-training median with whether the lagged 63-day mean score gap is negative. Both rules estimate one weight per state using the same past observations and the same weight bounds, $\operatorname{logistic}(-8)$ and $\operatorname{logistic}(8)$. StateDirect minimizes historical mixture FZ0 within each state, and StateGap applies the negative sigmoid to the state's mean normalized gap clipped to $[-8,8]$. A feature-free direct-score control estimates one constant weight on the same observations. All three update on the same monthly calendar.

\subsection{From Endpoint Scores to Mixture Decisions}\label{sec:surrogate}
The learner's surrogate target and the mixture decision solve different problems. Write $v_t(g)=v_{5,t}+g\delta v_t$, where $\delta v_t=v_{6,t}-v_{5,t}$, and hold the common tail constants $c_{q,t},c_{e,t}$ fixed at the forecast origin. For a candidate weight $g\in[0,1]$, define
\begin{align*}
 q_t(g)&=c_{q,t}\sqrt{v_t(g)},\qquad
 e_t(g)=c_{e,t}\sqrt{v_t(g)},\\
 S_t(g)&:=S_t\bigl(q_t(g),e_t(g)\bigr).
\end{align*}
Let $\mathbf1\{A\}$ denote the indicator of event $A$. For $L_t\ne q_t(g)$, differentiating the realized score gives
\begin{equation*}
 \frac{\partial S_t(g)}{\partial g}
 =\frac{\delta v_t}{2v_t(g)}
 \left[1-\frac{L_t\mathbf1\{L_t>q_t(g)\}}{\alpha e_t(g)}\right].
\end{equation*}
Assume an integrable, continuous conditional loss law, positive endpoint variances, and positive forecast constants and tail probability. The population gap is
\begin{equation*}
 \mu_t:=\mathrm E[\Delta S_t\mid\mathcal F_{t-1}]
 =\int_0^1\mathrm E\!\left[\frac{\partial S_t(g)}{\partial g}
                  \,\middle|\,\mathcal F_{t-1}\right]dg.
\end{equation*}
The gap integrates the gradient over the weight interval. The optimizer depends on how the expected score changes along that interval.

\begin{proposition}[Population-Optimal Variance-Mixture Weight]\label{prop:population-optimizer}
Under the stated assumptions, write $\sigma=\sqrt{v_t(g)}$ and $\overline S_t(\sigma)=\mathrm E[S_t(c_{q,t}\sigma,c_{e,t}\sigma)\mid\mathcal F_{t-1}]$. On the positive scale interval between the two endpoints,
\begin{equation}\label{eq:scale-score-derivative}
 \frac{d\overline S_t(\sigma)}{d\sigma}
 =\frac{H_t(\sigma)}{\alpha c_{e,t}\sigma^2},\qquad
 H_t(\sigma)=\alpha c_{e,t}\sigma-
 \mathrm E[L_t\mathbf1\{L_t>c_{q,t}\sigma\}\mid\mathcal F_{t-1}].
\end{equation}
If the endpoint variances differ, let $\sigma_{\min}$ and $\sigma_{\max}$ be the lower and upper scales. The unique minimizing scale $\sigma_\star$ is $\sigma_{\min}$ if $H_t(\sigma_{\min})\geq 0$, $\sigma_{\max}$ if $H_t(\sigma_{\max})\leq 0$, and the unique interior root of $H_t$ otherwise. The minimizing weight is $g^\star=(\sigma_\star^2-v_{5,t})/\delta v_t$, including when the endpoint scales are in reverse order. If $\delta v_t=0$, all weights issue identical forecasts and attain the same score.
\end{proposition}

\begin{proof}
Appendix~\ref{app:positive_control} justifies differentiating under the conditional expectation. The expected truncated moment is continuous and nonincreasing in $\sigma$ because raising a positive threshold removes only positive loss contributions. Thus $H_t$ is continuous and strictly increasing. The denominator in Equation~\eqref{eq:scale-score-derivative} is positive, so the respective endpoint signs give the boundary minima. With opposite signs, the derivative changes from negative to positive at the unique bracketed root. When the variances differ, the affine variance map recovers the unique weight from the minimizing scale. Coincident variances give a constant forecast across all weights.
\end{proof}

Computing the root under the true conditional law gives an evaluation-only oracle.

\begin{proposition}[Non-Identification from Endpoint Scores]\label{prop:endpoint-nonidentification}
The two endpoint VaR--ES forecasts and their conditional expected FZ0 difference are in general insufficient to uniquely determine the optimal interior variance-mixture weight.
\end{proposition}

\begin{proof}
Set $\alpha=0.025$, use normal VaR--ES constants and fix endpoint scales at 0.75 and approximately 1.527526, with the upper endpoint chosen to give a zero gap under $N(0,1)$. A centered mixture of normals with standard deviations 0.6 and 1.5 and probability approximately 0.756092 on the narrower component has the same zero gap by exact interpolation of the component gaps. The unique optimal variance-mixture weights are approximately 0.247059 and 0.277607, respectively (Appendix~\ref{app:positive_control}). Any rule based only on the conditional gap and the common endpoint forecasts must return the same weight for these two laws and therefore cannot be optimal for both. The zero-gap sigmoid selects 0.5 in both cases, with expected FZ0 regret of approximately 0.067257 and 0.033070, respectively.
\end{proof}

Near zero, $g_t\simeq 0.5-f_m(X_{t-1})/4$. Predicted gaps that are small relative to the historical target standard deviation therefore produce weights close to 0.5. M13 tests the effect of increasing the weight's sensitivity to predicted gaps, while M14 and M16 compare training objectives within the same bounded linear family. Section~\ref{sec:positive_control} compares the exact-gap sigmoid with the continuous feasible optimum, with a 101-point grid used as a numerical check.

\section{Forecast Evaluation Framework}\label{sec:evaluation}
\subsection{Accuracy and Paired Uncertainty}
For realized loss $L_t$, positive issued VaR and ES forecasts $q,e$, and $\alpha=0.025$, the joint score is
\begin{equation}\label{eq:fz}
 S_t(q,e)=\frac{\mathbf1\{L_t>q\}(L_t-q)}{\alpha e}
 +\frac{q}{e}+\log(e)-1.
\end{equation}
The realized score for model $j$ is $S_t(q_{j,t},e_{j,t})$. Exceedance diagnostics use the strict inequality $L_t>q$, and the score's hinge term is zero at a tie.
Lower scores indicate better forecasts \citep{patton2019dynamic}. A common positive rescaling of losses and forecasts adds the same log constant to both scores, leaving paired differences unchanged. Comparisons use common dates within each portfolio. Unless stated otherwise, differences are target minus control, so negative values favor the target.

We construct basic 95\% intervals from 1,999 paired circular block bootstrap draws. For observed mean difference $\widehat d$ and resampled means $\widehat d_b^*$, the centered two-sided test reports $\{1+\sum_b\mathbf1(|\widehat d_b^*-\widehat d|\geq|\widehat d|)\}/2000$. The main comparisons use 20-day blocks. The designated P1 endpoint result also uses 10-day blocks, and the sensitivity tables use 10, 20 and 60 days. Resampling common chronological blocks maintains the forecast pairing. Inference conditions on the fitted forecasts and selected specifications.

We apply Holm's adjustment \citep{holm1979simple} separately to the five M7 contrasts within a portfolio in the endpoint analysis, the twelve M7-minus-M8 contrasts across portfolios and estimation windows, the four M14-minus-M16 contrasts, and the four portfolios for each supplementary DCC and state-rule contrast. Each block length defines a separate family for these comparisons. The full-estimation simulation has a separate family of eight tuned-M7-minus-M8 comparisons. Confidence intervals are pointwise.

We construct a 90\% model confidence set (MCS) for the initial candidate set of eleven models (M1--M11) using the main estimation window, the range statistic and ten-day circular blocks \citep{hansen2011model}. The additional specifications M12--M16 are evaluated separately through paired score comparisons. The MCS results therefore describe relative forecasting performance within the specified eleven-model candidate set. Appendix~\ref{app:mcs} reports membership by portfolio.

\subsection{Calibration}
For a given model in this subsection, write $q_t,e_t$ for its issued forecasts, suppressing the model subscript. Coverage, independence, conditional coverage and dynamic quantile (DQ) tests examine exceedance behavior. DQ uses an intercept, four lagged centered hits and current VaR. The joint identification-moment diagnostic tests the means of
\begin{equation*}
 m_{1,t}=\mathbf1\{L_t>q_t\}-\alpha,\quad
 m_{2,t}=\frac{\mathbf1\{L_t>q_t\}(L_t-q_t)/\alpha+q_t-e_t}{e_t},
\end{equation*}
and their products with centered standardized $\log(e_t)$. For each portfolio, estimation window and evaluated model separately, let $\mathcal T$ contain the 3,772 common evaluation dates in 2011--2025 and set $n=|\mathcal T|$. Define
\begin{align*}
 \bar\ell&:=\frac{1}{n}\sum_{t\in\mathcal T}\log(e_t),\qquad
 s_\ell^2:=\frac{1}{n}\sum_{t\in\mathcal T}
                  [\log(e_t)-\bar\ell]^2,\\
 \xi_t&:=\frac{\log(e_t)-\bar\ell}{s_\ell}.
\end{align*}
The four moments are $(m_{1,t},m_{2,t},m_{1,t}\xi_t,m_{2,t}\xi_t)^{\prime}$. Standardization uses the evaluation sample and variance divisor $n$ for this diagnostic only. A Wald test uses Bartlett heteroskedasticity- and autocorrelation-consistent (HAC) covariance with lag four \citep{newey1987simple}.

We also apply the strict and auxiliary expected shortfall regression (ESR) tests of \citet{bayer2022regression} using lower-tail inputs $(-L_t,-q_t,-e_t)$ and $\alpha=0.025$. Both test a zero ES intercept and a unit ES slope. Strict ESR uses the ES forecast in both equations; auxiliary ESR uses VaR in the quantile equation and ES in the ES equation. We use the misspecification-robust asymptotic covariance, without HAC correction or bootstrap resampling. The 168 tests use common evaluation dates. The twenty-four tests for M7 form a supplementary Holm family.

\section{Empirical Results}\label{sec:empirical}
\subsection{Learning Increment and Endpoint Gains}
The primary comparison is M7 versus equal mixing (M8). In P1 at $W=756$, the mean difference is approximately $-0.0017$ FZ0, with a 20-day pointwise 95\% confidence interval of $[-0.0035,\,0.0002]$ and a Holm-adjusted $p$-value of 0.576 across the twelve portfolio--window comparisons. The interval leaves open a small improvement or deterioration. Table~\ref{tab:adaptive_p1} shows that M6 and M12 have lower mean scores than M7 in P1. Table~\ref{tab:adaptive_pairs} compares M7 with the individual endpoints and simple mixture controls.
\begin{table}[!htbp]\centering
\caption{Main-window P1 forecasts on common 2011--2025 dates.}\label{tab:adaptive_p1}
\begingroup\small\singlespacing
\setlength{\tabcolsep}{4pt}
\begin{tabular}{lrrrr}\toprule
Method & Mean FZ0 & Hit rate (\%) & DQ $p$ & Joint $p$\\\midrule
M1 HS & $-3.972$ & 2.49 & $<0.001$ & 0.045\\
M2 EWMA-N & $-4.109$ & 3.29 & $<0.001$ & $<0.001$\\
M3 GARCH-N & $-4.175$ & 3.00 & $<0.001$ & 0.001\\
M4 GARCH-$t$ & $-4.183$ & 2.86 & 0.001 & 0.217\\
M5 EQ-$t$ & $-4.166$ & 3.13 & $<0.001$ & 0.088\\
M6 EW-$t$ & $-4.203$ & 2.68 & 0.105 & 0.195\\
M7 ML & $-4.199$ & 2.78 & 0.097 & 0.238\\
M8 Equal mixing & $-4.197$ & 2.84 & 0.074 & 0.262\\
M9 Volatility rule & $-4.172$ & 3.00 & $<0.001$ & 0.135\\
M10 Score rule & $-4.192$ & 2.81 & 0.065 & 0.134\\
M11 Expanding mean & $-4.197$ & 2.81 & 0.102 & 0.259\\
M12 Fixed weight & $-4.204$ & 2.62 & 0.090 & 0.180\\
\bottomrule\end{tabular}
\par\medskip\footnotesize\textit{Notes:} $N=3,772$, $W=756$. The nominal hit rate is 2.5\%. In model labels, N and $t$ denote normal and Student-$t$ innovations, and ML denotes machine learning. Joint denotes the four-moment HAC test. All diagnostic $p$-values are unadjusted.
\endgroup\end{table}
\begin{table}[!htbp]\centering
\caption{M7 increments over endpoint and simple-mixture controls.}\label{tab:adaptive_pairs}
\begingroup\small\singlespacing
\setlength{\tabcolsep}{4pt}
\begin{tabular}{llrr}\toprule
Portfolio & Control & M7 minus control & 95\% interval\\\midrule
P1 & M8 Equal mixing & $-1.718$ & \tableinterval[99]{-3.482}{0.167}\\
P1 & M5 EQ-$t$ & $-32.902$ & \tableinterval[99]{-53.754}{-10.594}\\
P1 & M6 EW-$t$ & 4.106 & \tableinterval[99]{-12.916}{20.394}\\
P1 & M11 Expanding mean & $-1.631$ & \tableinterval[99]{-3.275}{0.164}\\
P1 & M12 Fixed weight & 5.230 & \tableinterval[99]{-8.281}{18.254}\\
P2 & M8 Equal mixing & $-0.029$ & \tableinterval[99]{-1.218}{1.452}\\
P2 & M5 EQ-$t$ & $-13.268$ & \tableinterval[99]{-28.815}{3.565}\\
P2 & M6 EW-$t$ & $-4.542$ & \tableinterval[99]{-20.387}{12.115}\\
P2 & M11 Expanding mean & $-0.167$ & \tableinterval[99]{-1.319}{1.332}\\
P2 & M12 Fixed weight & $-7.019$ & \tableinterval[99]{-17.435}{4.319}\\
P3 & M8 Equal mixing & $-0.116$ & \tableinterval[99]{-0.863}{0.905}\\
P3 & M5 EQ-$t$ & $-5.667$ & \tableinterval[99]{-18.704}{8.863}\\
P3 & M6 EW-$t$ & $-6.452$ & \tableinterval[99]{-20.145}{7.850}\\
P3 & M11 Expanding mean & $-0.223$ & \tableinterval[99]{-0.923}{0.761}\\
P3 & M12 Fixed weight & $-5.667$ & \tableinterval[99]{-18.704}{8.863}\\
P4 & M8 Equal mixing & 0.318 & \tableinterval[99]{-0.332}{1.018}\\
P4 & M5 EQ-$t$ & 2.498 & \tableinterval[99]{-7.485}{12.709}\\
P4 & M6 EW-$t$ & $-11.608$ & \tableinterval[99]{-22.793}{-0.219}\\
P4 & M11 Expanding mean & 0.327 & \tableinterval[99]{-0.285}{0.972}\\
P4 & M12 Fixed weight & 2.498 & \tableinterval[99]{-7.485}{12.709}\\
\bottomrule\end{tabular}
\par\medskip\footnotesize\textit{Notes:} Differences and interval endpoints are reported in units of $10^{-3}$ FZ0. $W=756$, $N=3,772$. Pointwise basic intervals use 1,999 paired bootstrap draws with 20-day blocks.
\endgroup\end{table}

For P1, M7 lowers mean FZ0 by approximately 0.0329 relative to M5. The basic 95\% interval with ten-day blocks is $[-0.0537,\,-0.0105]$. Within the five-comparison P1 family, the Holm-adjusted $p$-values are 0.023 and 0.015 for ten- and twenty-day blocks, respectively.

For each portfolio, the pointwise intervals against M11 (updating with no state features) and M12 (fixed after validation) include zero. These comparisons leave uncertainty about the value of state inputs and time variation in the weights. The P4 interval against the exponentially weighted endpoint lies below zero. Both M7 and M8 are in the main-window MCS for all four portfolios.

\subsection{Window Sensitivity}
Table~\ref{tab:adaptive_windows} compares M7 with equal mixing across four portfolios and three estimation windows, with separate validation for each window. The pointwise interval favors M7 for P1 at 1,008 days and M8 for P3 at 504 days. The twelve-comparison Holm adjustment yields no rejections. Appendix Table~\ref{tab:window_summary} reports sensitivity to the block length.
\begin{table}[!htbp]\centering
\caption{Primary simple-mixture comparison across estimation windows.}\label{tab:adaptive_windows}
\begingroup\small\singlespacing
\setlength{\tabcolsep}{4pt}
\begin{tabular}{lrrrr}\toprule
Portfolio & $W$ & M7 minus M8 & 95\% interval & Holm $p$\\\midrule
P1 & 504 & 0.348 & \tableinterval{-0.840}{1.464} & 1.000\\
P1 & 756 & $-1.718$ & \tableinterval{-3.482}{0.167} & 0.576\\
P1 & 1,008 & $-0.941$ & \tableinterval{-1.693}{-0.175} & 0.210\\
P2 & 504 & 0.228 & \tableinterval{-1.172}{1.660} & 1.000\\
P2 & 756 & $-0.029$ & \tableinterval{-1.218}{1.452} & 1.000\\
P2 & 1,008 & 0.364 & \tableinterval{-0.006}{0.687} & 0.450\\
P3 & 504 & 0.875 & \tableinterval{0.084}{1.628} & 0.308\\
P3 & 756 & $-0.116$ & \tableinterval{-0.863}{0.905} & 1.000\\
P3 & 1,008 & 0.063 & \tableinterval{-0.349}{0.482} & 1.000\\
P4 & 504 & 0.623 & \tableinterval{-0.675}{2.151} & 1.000\\
P4 & 756 & 0.318 & \tableinterval{-0.332}{1.018} & 1.000\\
P4 & 1,008 & 0.148 & \tableinterval{-0.825}{1.247} & 1.000\\
\bottomrule\end{tabular}
\par\medskip\footnotesize\textit{Notes:} Differences and interval endpoints are reported in units of $10^{-3}$ FZ0. Negative differences favor M7. All windows use common dates. Intervals are pointwise, and Holm $p$-values cover the twelve comparisons with 20-day blocks.
\endgroup\end{table}

\subsection{Signal, Weight Amplification and Direct Training Objectives}\label{sec:mechanism}
The diagnostics distinguish three potential limits on the value of state information (Section~\ref{sec:surrogate}): the accuracy of the predicted endpoint-score gap, the sensitivity of mixture weights to that prediction, and the choice of training objective. Table~\ref{tab:adaptive_signal} addresses the first question by comparing M7's predictions with an expanding historical mean updated on the same dates. All four incremental $R^2$ estimates are negative, and their 20-day intervals and the HAC intervals for the incremental-regression slopes include zero. The prediction exercise therefore gives little evidence of a signal beyond the historical mean.
\begin{table}[!htbp]\centering
\caption{Endpoint-gap prediction, M7 mixture weights and validation choices.}\label{tab:adaptive_signal}
\begingroup\small\singlespacing
\setlength{\tabcolsep}{4pt}
\begin{tabular}{lrrrrrr}\toprule
Portfolio & $10^3 R^2_{\rm inc}$ & 95\% interval & Mean $g$ & SD of $g$ & M12 weight & M13 $k$\\\midrule
P1 & $-4.970$ & \tableinterval[99]{-16.770}{9.030} & 0.508 & 0.031 & 0.900 & 0\\
P2 & $-1.500$ & \tableinterval[99]{-12.420}{9.920} & 0.501 & 0.018 & 0.150 & 4\\
P3 & $-0.740$ & \tableinterval[99]{-8.730}{7.380} & 0.496 & 0.018 & 0.000 & 4\\
P4 & $-8.560$ & \tableinterval[99]{-16.630}{3.730} & 0.490 & 0.025 & 0.000 & 0\\
\bottomrule\end{tabular}
\par\medskip\footnotesize\textit{Notes:} The $10^3$ multiplier applies only to incremental $R^2$ and its interval endpoints. The weight summaries and the mean path-level conditional-gap $R^2$ in Appendix Tables~\ref{tab:matched_diagnostics} and \ref{tab:positive_diagnostics} are reported on their original scales. For predictions of the realized endpoint-score gap, $R^2_{\rm inc}=1-\mathrm{MSE}_{7}/\mathrm{MSE}_{\rm mean}$, where MSE denotes mean squared error. Intervals use 20-day blocks. Weight means and standard deviations (SDs) use evaluation dates. The M12 weight and M13 amplification coefficient $k$ are selected in 2010.
\endgroup\end{table}

\begin{figure}[!htbp]\centering
\includegraphics[width=\linewidth]{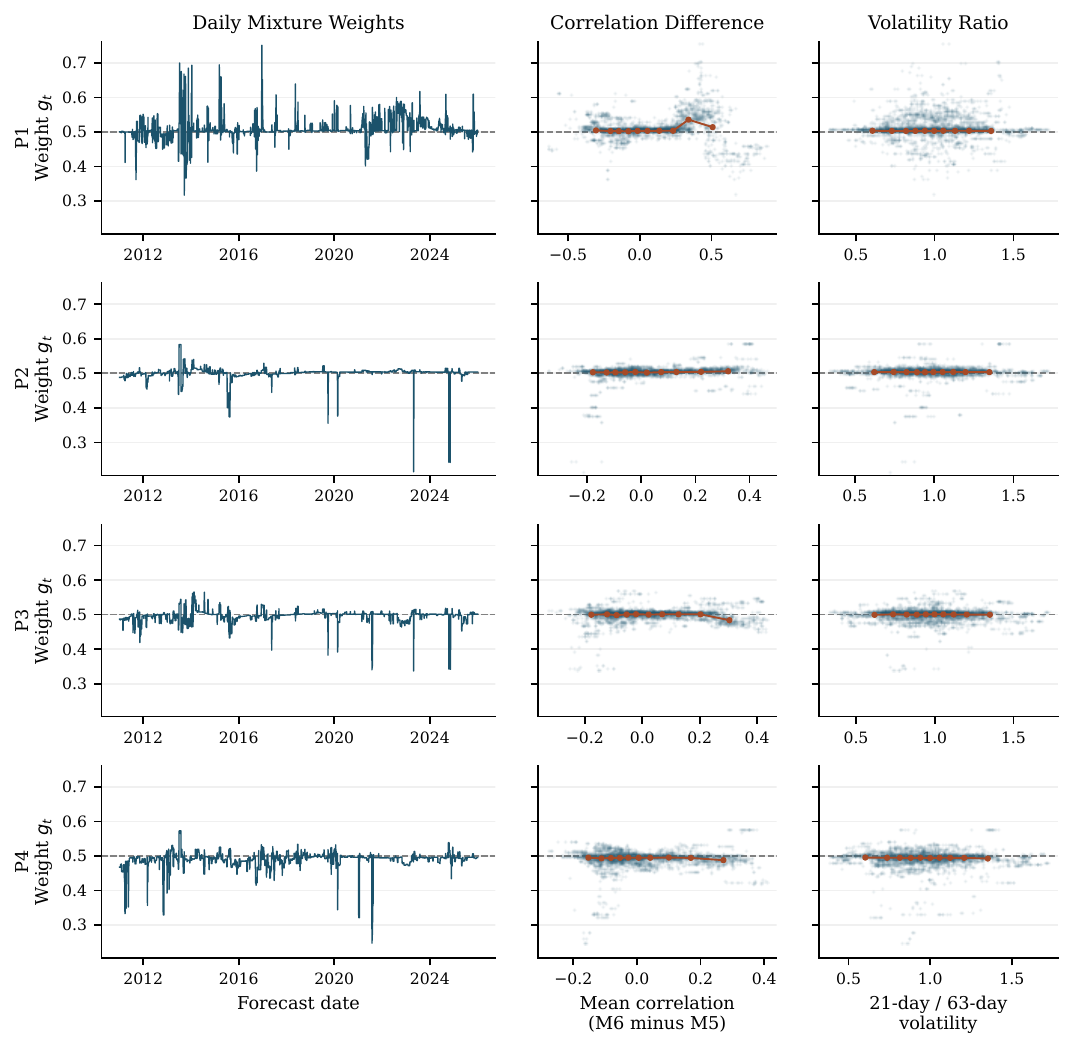}
\caption{M7 mixture weights on the exponentially weighted correlation estimate (M6). Left: daily weights. Center: weights against the difference in mean off-diagonal correlations (M6 minus M5). Right: weights against the lagged ratio of 21-day to 63-day portfolio volatility. Dashed lines mark equal mixing. Orange lines connect the medians within ten quantile bins.}\label{fig:gate_mechanism}
\end{figure}

Figure~\ref{fig:gate_mechanism} shows how these predictions translate into weights. The binned medians remain near 0.5 despite occasional larger departures, and the sample standard deviation of the weights ranges from 0.018 to 0.031 at the main window. This motivates checking whether a larger response to the predictor improves the score.

M13 probes that response by varying the amplification of the feature-based component. Validation removes it for P1 and P4 and selects a coefficient of four for P2 and P3. All four intervals against M7 include zero (Table~\ref{tab:gain_summary}). Supplementary state-conditional comparisons provide limited evidence that the average results conceal systematic gains from state-dependent weighting (Appendix~\ref{app:calibration_summary}).

Amplification retains the original prediction objective. To test whether that objective limits performance, M14 targets the mixture score directly, while M16 predicts the endpoint-score gap within the same bounded linear sigmoid family. Their intervals include zero at each of the three block lengths (Tables~\ref{tab:adaptive_direct} and \ref{tab:direct_summary}). At the main block length, the same holds for comparisons of M14 with M7, equal mixing, the expanding-mean gate, validation-selected fixed mixing, and the monthly constant-weight control M15. M14 weights vary considerably for P1 and P2, but closely track M15 for P3 and P4. Together, the diagnostics leave open how much of the limited incremental gain comes from signal quality or weight selection.
\begin{table}[!htbp]\centering
\caption{Direct-score training versus matched linear score-gap training.}\label{tab:adaptive_direct}
\begingroup\small\singlespacing
\setlength{\tabcolsep}{4pt}
\begin{tabular}{lrrr}\toprule
Portfolio & M14 minus M16 & 95\% interval & Holm $p$\\\midrule
P1 & 5.477 & \tableinterval[99]{-10.175}{19.586} & 1.000\\
P2 & 4.995 & \tableinterval[99]{-4.990}{15.488} & 1.000\\
P3 & 6.648 & \tableinterval[99]{-1.984}{14.992} & 0.486\\
P4 & $-0.309$ & \tableinterval[99]{-9.110}{8.951} & 1.000\\
\bottomrule\end{tabular}
\par\medskip\footnotesize\textit{Notes:} Differences and interval endpoints are reported in units of $10^{-3}$ FZ0. M14 and M16 share a bounded linear sigmoid family and select penalties separately in 2010. Intervals use 20-day blocks. Holm $p$-values cover the four portfolio comparisons.
\endgroup\end{table}

DCC lowers P1 FZ0 by approximately 0.0304 relative to M5 (Holm $p=0.042$ at 20 days). Its intervals relative to equal mixing and M7 include zero for all four portfolios. This supports distinguishing gains over one correlation estimator from gains over simple mixing.

In the matched state comparison, the differences have positive point estimates in P1--P3 and a slightly negative estimate in P4 (Table~\ref{tab:new_dependence}). None rejects after the four-portfolio Holm adjustment. The intervals relative to the re-estimated feature-free direct-score weight also include zero in each portfolio. Appendix Table~\ref{tab:dependence_supplement} reports both the DCC-versus-M7 and the StateDirect-versus-feature-free contrasts. Direct training increases weight dispersion substantially, with standard deviations of 0.273--0.372 compared with 0.008--0.012 for the gap-based rule.

\begin{table}[!htbp]\centering\footnotesize\singlespacing
\setlength{\tabcolsep}{4pt}
\caption{DCC benchmarks and matched state rules at the main window.}\label{tab:new_dependence}
\begin{tabular}{lrrrrr}\toprule
Portfolio & DCC--M5 & DCC--M8 & \shortstack{StateDirect\\minus StateGap} & \shortstack{Direct\\weight SD} & \shortstack{Gap\\weight SD}\\\midrule
P1 & \shortstack{$-30.413$\\{\tableinterval[99]{-53.945}{-5.448}}\\0.042} & \shortstack{0.771\\{\tableinterval{-6.947}{8.477}}\\1.000} & \shortstack{4.061\\{\tableinterval[99]{-7.017}{15.902}}\\0.976} & 0.273 & 0.008\\
P2 & \shortstack{$-13.937$\\{\tableinterval[99]{-28.869}{0.992}}\\0.216} & \shortstack{$-0.699$\\{\tableinterval{-5.836}{4.618}}\\1.000} & \shortstack{8.851\\{\tableinterval[99]{0.106}{16.910}}\\0.156} & 0.318 & 0.009\\
P3 & \shortstack{$-6.578$\\{\tableinterval[99]{-19.308}{5.929}}\\0.636} & \shortstack{$-1.027$\\{\tableinterval{-4.898}{3.193}}\\1.000} & \shortstack{6.714\\{\tableinterval[99]{-4.782}{17.904}}\\0.708} & 0.372 & 0.012\\
P4 & \shortstack{1.119\\{\tableinterval[99]{-8.533}{10.452}}\\0.833} & \shortstack{$-1.062$\\{\tableinterval{-4.162}{2.628}}\\1.000} & \shortstack{$-0.196$\\{\tableinterval[99]{-8.487}{7.892}}\\0.976} & 0.304 & 0.009\\
\bottomrule\end{tabular}
\par\smallskip\begin{minipage}{0.98\linewidth}\scriptsize\textit{Notes:} Each contrast cell gives the mean difference, pointwise 95\% interval and Holm $p$-value, in that order. Scores and endpoints are in $10^{-3}$ FZ0. The block length is 20 days, and each contrast is adjusted across the four portfolios. Weight standard deviations are unscaled.\end{minipage}
\end{table}

\subsection{Calibration Findings}
Main-window M7 exceedance rates range from 2.65 to 2.78\% (Table~\ref{tab:adaptive_calibration}). The joint HAC moment test rejects at 5\% for P3 and P4. Across the three windows, six of the twenty-four strict and auxiliary ESR tests for M7 reject at the pointwise 5\% level, with zero rejections after Holm adjustment. Some direct-objective forecasts also yield DQ or joint-moment rejections. Appendix~\ref{app:calibration_summary} reports the remaining ESR results.
\begin{table}[!htbp]\centering
\caption{M7 calibration at the main window.}\label{tab:adaptive_calibration}
\begingroup\small\singlespacing
\setlength{\tabcolsep}{4pt}
\begin{tabular}{lrrrrr}\toprule
Portfolio & Hit rate (\%) & DQ $p$ & Joint $p$ & Strict ESR $p$ & Aux. ESR $p$\\\midrule
P1 & 2.78 & 0.097 & 0.238 & 0.059 & 0.062\\
P2 & 2.78 & 0.303 & 0.205 & 0.061 & 0.067\\
P3 & 2.68 & 0.641 & 0.035 & 0.570 & 0.058\\
P4 & 2.65 & 0.200 & 0.015 & 0.128 & 0.060\\
\bottomrule\end{tabular}
\par\medskip\footnotesize\textit{Notes:} All tests use the same 3,772 dates. Joint uses a Bartlett HAC covariance. ESR uses a misspecification-robust asymptotic covariance. Reported $p$-values are unadjusted.
\endgroup\end{table}

\section{Supplementary Economic Evaluation}\label{sec:economics}
We assess hypothetical ES-based buffer policies on common evaluation dates. The normalized cost combines the mean buffer with the mean uncovered loss using illustrative carrying-cost ratios. The validation-budget rule equalizes mean buffers during 2010, and mean buffers can differ during evaluation. Appendix~\ref{app:economic_sim_summary} reports the policy definitions and exploratory cost comparisons.

\section{Monte Carlo Simulation}\label{sec:simulation}
\subsection{Data-Generating Processes and Forecast Design}
We simulate fifty independent paths for each combination of stable or switching correlation, normal or Student-$t_5$ innovations, and two or eight nested assets. Within-class innovation correlation is 0.8. Cross-class correlation is $0.8\rho_t$, where $\rho_t$ is the correlation between two common Gaussian factors. Stable designs set $\rho_t=-0.3$. Switching designs alternate $-0.3,0.6,-0.3,0.6$ in four consecutive 250-day forecast blocks. The Gaussian innovations combine the common factors with idiosyncratic normals. For Student tails, multiply the innovation vector at each date by $\sqrt{3/Q_t}$, where the $Q_t$ are independent and identically distributed (i.i.d.) as $\chi^2_5$ and independent of the Gaussian draws. The common scaling preserves unit marginal innovation variance and gives elliptical tails.

Marginal conditional variances follow
\begin{equation*}
 h_{i,t+1}=0.04\overline h_i+0.06r_{i,t}^2+0.90h_{i,t}.
\end{equation*}
Here $\overline h_i$ is the unconditional marginal variance, equal to $0.012^2$ for first-class assets and $0.007^2$ for second-class assets. The two-asset weights are $(0.6,0.4)$, and the eight-asset weights are 0.15 for each first-class asset and 0.10 for each second-class asset. Scenarios and dimensions share random draws within a path. A 1,000-day burn-in and a 756-day estimation history precede the 1,000 forecast dates. M5--M7 estimate the marginal models and common predictive tail parameter.

Forecast days 0--249 initialize learning, days 250--499 provide prequential validation among the four tree candidates, and days 500--999 provide evaluation. Expanding refits occur every 21 days and restart on that schedule at day 500. The 63-day score-history requirement leaves 187 eligible initial observations and 437 at the start of evaluation. We compare tuned M7 with a fixed tree specification refitted on the same dates and with the non-ML controls. The fixed tree specification uses a maximum depth of one and 40 trees. All forecasting and selection use only each path's observed history.

\subsection{Scores, Error Metrics and Replication-Level Inference}\label{subsec:simulation-error-metrics}
For each method and path, realized FZ0 is averaged over the 500 evaluation dates. Let $e_t^{\rm oracle}=e_t^{\ast}$ denote the conditional ES under the true data-generating process, computed analytically from its marginal variance, correlation and tail law. Suppressing the model subscript, let $e_t$ be the issued ES forecast being evaluated. The secondary relative root mean squared error (RRMSE) is
\begin{equation*}
 \operatorname{RRMSE}(e)=\left\{\frac1{500}\sum_{t=500}^{999}
       (e_t/e_t^{\rm oracle}-1)^2\right\}^{1/2}.
\end{equation*}
We use oracle ES only to measure forecast error. Paired path-level differences across fifty independent paths give Monte Carlo standard errors (MCSEs) and $t_{49}$ intervals. Holm adjustment covers eight full-period tuned-M7-minus-M8 FZ0 comparisons. Appendix Table~\ref{tab:sim_periods} reports separate results for days 500--749 and 750--999.

\begin{figure}[!htbp]\centering
\includegraphics[width=\linewidth]{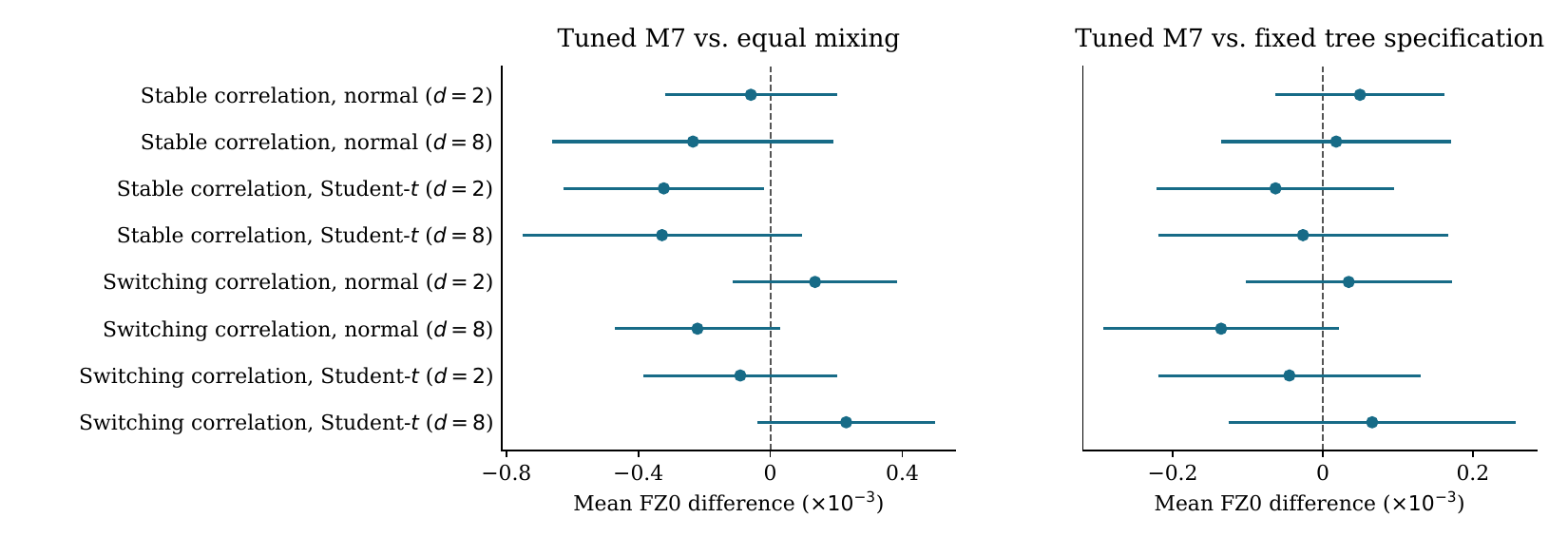}
\caption{Mean differences in the joint Fissler--Ziegel score (FZ0) over the 500 evaluation dates for tuned M7 versus equal mixing (left) and versus the fixed tree specification refitted on the same dates (right). The fixed specification uses depth one and 40 trees. Points and pointwise 95\% $t_{49}$ intervals use fifty paired paths. Negative differences favor tuned M7.}\label{fig:adaptive_sim}
\end{figure}

\begin{table}[!htbp]\centering
\caption{Monte Carlo simulation results: tuned M7 versus equal mixing.}\label{tab:adaptive_sim}
\begingroup\small\singlespacing
\setlength{\tabcolsep}{4pt}
\begin{tabular}{lrrrrrr}\toprule
Mechanism & $d$ & FZ0 diff. & FZ0 interval & Holm $p$ & ES error diff. & ES error interval\\\midrule
Stable N & 2 & $-0.059$ & \tableinterval{-0.320}{0.202} & 1.000 & $-0.490$ & \tableinterval{-0.788}{-0.191}\\
Stable N & 8 & $-0.235$ & \tableinterval{-0.661}{0.192} & 1.000 & $-0.678$ & \tableinterval{-1.010}{-0.347}\\
Stable $t$ & 2 & $-0.323$ & \tableinterval{-0.628}{-0.018} & 0.306 & $-0.570$ & \tableinterval{-0.949}{-0.191}\\
Stable $t$ & 8 & $-0.328$ & \tableinterval{-0.751}{0.095} & 0.627 & $-0.552$ & \tableinterval{-0.848}{-0.255}\\
Switching N & 2 & 0.135 & \tableinterval{-0.114}{0.385} & 1.000 & $-0.242$ & \tableinterval{-0.396}{-0.089}\\
Switching N & 8 & $-0.221$ & \tableinterval{-0.471}{0.029} & 0.573 & $-0.205$ & \tableinterval{-0.362}{-0.048}\\
Switching $t$ & 2 & $-0.091$ & \tableinterval{-0.384}{0.201} & 1.000 & 0.099 & \tableinterval{-0.097}{0.294}\\
Switching $t$ & 8 & 0.230 & \tableinterval{-0.039}{0.499} & 0.573 & $-0.058$ & \tableinterval{-0.280}{0.164}\\
\bottomrule\end{tabular}
\par\medskip\footnotesize\textit{Notes:} Differences and interval endpoints for FZ0 are reported in units of $10^{-3}$ FZ0. Fifty paired paths, each with 500 evaluation dates. Intervals use $t_{49}$ and replication-level MCSEs. Holm $p$-values cover full-period FZ0 contrasts. ES RRMSE differences are changes in dimensionless ratios. They and their interval endpoints are scaled by $10^3$. Their intervals are secondary and unadjusted. N and $t$ denote normal and standardized $t_5$ innovations.
\endgroup\end{table}
Table~\ref{tab:adaptive_sim} and Figure~\ref{fig:adaptive_sim} provide little evidence that tuned M7 improves joint VaR--ES forecast accuracy over equal mixing. The pointwise interval favors M7 in the stable two-asset Student-$t$ design. None of the full-period FZ0 comparisons with equal mixing is significant after Holm adjustment. Comparisons with the expanding-mean gate and fixed tree specification are also inconclusive. Under switching correlation, pointwise intervals favor M7 over validation-selected fixed mixing in all four designs, suggesting a possible benefit of adaptive weighting when dependence varies.

Six of eight pointwise intervals favor M7 over equal mixing for oracle-relative ES error. Evidence against the expanding-mean gate is mixed, and comparisons with the fixed tree specification are inconclusive. These secondary, unadjusted comparisons provide clearer evidence of gains in ES estimation than in joint VaR--ES forecast accuracy.

\subsection{Positive Control with Matched State Occupancy}\label{sec:positive_control}
The known-margin simulation uses two assets with marginal standard deviations $(0.012,0.007)$ and portfolio weights $(0.6,0.4)$. Innovations are normal or variance-standardized Student-$t_5$. Before the date-$t$ loss is realized, the forecaster observes an exogenous driver $X_t$ taking values $-1$ and $1$ with equal probability. It is independent across dates and independent of the Gaussian and chi-square draws and state uniforms. The forecast information set is $\mathcal G_t=\mathcal F_{t-1}\vee\sigma(X_t)$. For a uniform $U_t$,
\begin{equation*}
 p_t=\operatorname{logistic}(\beta X_t),\qquad
 \rho_t=-0.6+1.4\,\mathbf1\{U_t<p_t\},\qquad \beta\in\{0,3\}.
\end{equation*}
Here $p_t=\Pr(\rho_t=0.8\mid\mathcal G_t)$. At $\beta=0$, $p_t=0.5$. At $\beta=3$, it is approximately 0.0474 or 0.9526. Symmetry fixes the unconditional high-state probability at 0.5 in both designs. The endpoint forecasts share the normal or standardized Student-$t_5$ VaR--ES constants appropriate to the innovation law. Conditional portfolio losses are mixtures over the two correlation states.

Each of the 128 independent paths shares the driver, shocks and state uniforms across the four law--signal cells. Each cell has 1,000 evaluation dates and the same 756-day endpoint window and expanding 21-day refit schedule. In both signal designs, the gap learner uses the twelve features defined in Section~\ref{sec:gate} together with the observed driver. The boosted-tree specification is fixed at a maximum depth of one and 40 trees. A supplemental ridge learner includes interactions between the observed driver and the signed endpoint variance difference on the log scale. Latent states and true probabilities enter evaluation only. Appendix~\ref{app:positive_control} gives the fitting and evaluation details.

We report path-paired $t_{127}$ intervals and Monte Carlo standard errors. The change in the learner-minus-equal-mixture difference from $\beta=0$ to $\beta=3$ measures the incremental score effect of stronger state information. We adjust the two law-specific interaction tests for each learner separately. The four cell-level realized comparisons for the stump learner form their own Holm family.

\begin{table}[!htbp]\centering\footnotesize\singlespacing
\setlength{\tabcolsep}{4pt}
\caption{Effect of stronger state information on realized FZ0 relative to equal mixing.}\label{tab:matched_interaction}
\begin{tabular}{lrrrrrr}\toprule
Law & Stump change & MCSE & Holm $p$ & Ridge change & MCSE & Holm $p$\\\midrule
Normal & \shortstack{0.082\\{\tableinterval{-0.080}{0.245}}} & 0.082 & 0.319 & \shortstack{$-1.813$\\{\tableinterval{-2.243}{-1.384}}} & 0.217 & $<0.001$\\
$t_5$ & \shortstack{0.275\\{\tableinterval{0.064}{0.486}}} & 0.107 & 0.022 & \shortstack{$-0.895$\\{\tableinterval{-1.438}{-0.352}}} & 0.274 & 0.001\\
\bottomrule\end{tabular}
\par\smallskip\begin{minipage}{0.98\linewidth}\scriptsize\textit{Notes:} Changes compare $\beta=3$ with $\beta=0$. Negative values indicate a greater gain over equal mixing under strong state information. Changes, pointwise 95\% intervals, and Monte Carlo standard errors are in $10^{-3}$ FZ0. Each learner has a separate two-law Holm family.\end{minipage}
\end{table}

The high-state frequencies average 50.25\% and 50.11\% across paths. The stump learner's mean path-level gap $R^2$ relative to the expanding historical mean increases from $-0.462$ to $-0.059$ under normal innovations and from $-1.308$ to $-0.172$ under Student-$t_5$. Both the stump learner and the historical-mean benchmark have larger absolute gap MSEs when the driver is informative (Appendix Table~\ref{tab:matched_diagnostics}). The realized-score change is inconclusive under normal innovations and adverse under Student-$t_5$ (Table~\ref{tab:matched_interaction}). For the boosted-stump learner, the expected-score interaction intervals include zero under both innovation laws. The ridge learner with driver interactions improves realized scores relative to equal mixing in both strong-signal cells. A paired comparison with the same ridge specification restricted to omit the driver interactions corroborates the realized-score gains in these cells.

The \emph{learned-gap sigmoid} maps the stump learner's fitted gap to a weight. The \emph{exact-gap sigmoid} substitutes the true conditional gap and uses the same sigmoid and historical target scale. Their expected score difference can have either sign. Three evaluation-only oracles use the true conditional loss law. The \emph{feasible oracle} minimizes expected FZ0 over the endpoint scale interval using Proposition~\ref{prop:population-optimizer}. The \emph{ray oracle} retains the same VaR--ES ratio and allows any positive scale. The \emph{unrestricted VaR--ES oracle} issues the true conditional quantile and expected shortfall.

At $\beta=3$, the exact-gap sigmoid's expected regret relative to the feasible oracle is approximately 0.0216 under normal innovations and 0.0187 under Student-$t_5$ innovations. The learned-gap sigmoid incurs expected regret of approximately 0.0243 and 0.0207 in those cells. Boundary optima occur on 99.89\% and 98.94\% of dates, respectively. The feasibility loss is the expected FZ0 difference between the feasible and unrestricted oracles. The ray oracle isolates the contribution of the endpoint scale interval, which accounts for over 99.8\% of that loss under both laws when the driver is informative. Appendix Table~\ref{tab:matched_regret} reports paired intervals for mapping regret, learner regret and the total feasibility loss. Appendix Table~\ref{tab:matched_learner_half} reports each learner's score differences relative to equal mixing at both information strengths.

\section{Computational Cost and Limitations}\label{sec:robustness}

\subsection{Computational Cost}\label{sec:computation}
Table~\ref{tab:adaptive_cost} separates the cost of the shared endpoint forecasts from the incremental cost of M7. Computations were performed on a Windows 11 system with a four-core Intel processor, eight logical processors and approximately 4 gibibytes of memory, using Python 3.13.5, XGBoost 3.4.1 and \texttt{arch} 8.0.0. Numerical libraries and model fitting were restricted to a single thread. Total computational costs were obtained by aggregating the separately timed daily and monthly components over the evaluation period. Validation search was timed separately.
\begin{table}[!htbp]\centering
 \caption{Computational cost over the empirical evaluation schedule.}\label{tab:adaptive_cost}
\begingroup\small\singlespacing
\setlength{\tabcolsep}{4pt}
\begin{tabular}{lrrrrr}\toprule
Portfolio & Common base (s) & M7 increment (s) & M7 (ms/day) & M8 (ms/day) & Ratio\\\midrule
P1 & 28.8 & 124.5 & 40.7 & 7.7 & 5.31\\
P2 & 63.6 & 120.9 & 48.9 & 16.9 & 2.90\\
P3 & 111.6 & 102.4 & 56.7 & 29.6 & 1.92\\
P4 & 231.3 & 143.8 & 99.4 & 61.4 & 1.62\\
\bottomrule\end{tabular}
\par\medskip\footnotesize\textit{Notes:} Each portfolio has 3,772 evaluation dates and 180 monthly refits. Per-day costs amortize the timed components, including refits. The ratio compares M7 with M8.
\endgroup\end{table}

Across portfolios, M7 costs 1.62 to 5.31 times as much as M8 on an amortized basis. Training accounts for approximately 79--87\% of M7's incremental cost. Appendix~\ref{app:computation} reports component and validation timings.

\subsection{Limitations}
The historical evaluation period had already been examined, and the mechanism analyses reuse that period. The paired bootstrap conditions on fitted forecasts, selected specifications and estimated buffer multipliers. Its intervals therefore reflect variation in evaluation outcomes conditional on these choices. The score comparisons quantify uncertainty about differences in performance. Establishing equivalence would require an explicit margin and a corresponding test. Calibration tests assess specific restrictions on forecasts, and MCS membership describes relative score uncertainty within the candidate set.

The empirical scope is one-day VaR and ES at the 97.5\% confidence level for daily rebalanced gross portfolios of surviving ETFs. Provider-adjusted prices may be revised. The full-estimation simulation has a shorter learning history than the empirical evaluation. The matched-occupancy design isolates a strong, exogenous predictor in a two-asset model with known margins. Its gains arise under those information and model conditions. Independent evaluation would help assess the stability of the empirical findings.

\section{Conclusion}\label{sec:conclusion}
The retrospective M7-versus-M8 comparisons leave the incremental value of learning uncertain. For the stock--bond portfolio, both M7 and DCC improve on the equally weighted correlation estimator. The comparisons with equal mixing, the updating feature-free gate and the fixed validation weight leave open gains or losses from using state information and time-varying weights. Direct-score training generates more dispersed state weights, and its score advantage remains uncertain after adjustment.

Two conditional loss laws with identical endpoint forecasts and expected score gaps can require different interior weights. The matched-occupancy simulation shows how the resulting score gains depend on the learner's treatment of the state signal and the mapping from the signal to a weight. When information is strong, the ridge rule with driver interactions gains more from the signal, and the exact-gap sigmoid remains well above the feasible score optimum. These results motivate evaluating the weight decision against equal mixing and measuring the loss from the weight rule even when the surrogate signal is accurate.

\clearpage
\bibliography{paper}
\bibliographystyle{apalike}

\clearpage
\appendix
\counterwithin{table}{section}
\counterwithin{figure}{section}
\section{Learning Specifications and Diagnostics}\label{app:implementation}
The historical correlation features use normalized uncentered second moments. Matrix disagreements are root-mean-square off-diagonal differences. The 21- and 252-day residual windows and the preceding 21-day block are drawn from the trailing estimation history.
\begin{table}[!htbp]\centering
\caption{Validation-selected main-window M7 specifications.}\label{tab:adaptive_candidates}
\begingroup\small\singlespacing
\setlength{\tabcolsep}{4pt}
\begin{tabular}{lrrr}\toprule
Portfolio & Maximum depth & Number of trees & Learning rate\\\midrule
P1 & 2 & 80 & 0.03\\
P2 & 1 & 80 & 0.03\\
P3 & 2 & 40 & 0.03\\
P4 & 2 & 40 & 0.03\\
\bottomrule\end{tabular}
\par\medskip\footnotesize\textit{Notes:} Specifications minimize 2010 prequential FZ0.
\endgroup\end{table}

For the signal diagnostic, we recover the raw gap prediction by inverting the mixture sigmoid and multiplying by the training sample standard deviation of the target. The HAC incremental regression subtracts the expanding historical mean gap from both the realized and predicted gaps.

M14 and M16 standardize features using monthly training-sample means and sample standard deviations, then clip them to $[-5,5]$. Both use intercept bounds $[-8,8]$, slope bounds $[-3,3]$ and slope-only L2 penalties $\lambda_{\rm reg}\in\{0.001,0.01,0.1,1\}$. Each selects its penalty by unpenalized prequential FZ0 in 2010. We minimize the nonsmooth, nonconvex M14 objective using the limited-memory Broyden--Fletcher--Goldfarb--Shanno algorithm with bounds (L-BFGS-B) from multiple starting points. M16 solves a bounded quadratic score-gap problem.

M15 evaluates $g=0,0.01,\ldots,1$ and refines local grid minima by bounded scalar optimization, retaining both endpoints as candidates. The risk forecasts use the same variance mixture and tail constants as M7.

\paragraph*{Multistart Optimization.}
All 768 optimization runs initiated during validation and 2,880 runs initiated during evaluation converged for M14. The optimizer minimizes mean training FZ0 plus the slope-only L2 penalty. The prequential validation and evaluation forecast scores are unpenalized FZ0. The selected solution improved on the best initial evaluation objective by a median of $0.003625$ in the penalized objective. Terminal penalized objectives across starts within an evaluation fit differed by up to $0.008759$. Across the 192 validation fits, the selected penalized objective had median $-3.709980$ and range $[-3.980539,\,-3.329028]$. Across the 720 evaluation fits, its median was $-4.098183$ and its range was $[-4.332030,\,-3.500007]$.

\paragraph*{Scalar Optimization.}
The matched state fits and the feature-free direct-score control search 1,001 weights and refine candidate minima. The January fits also include the best 4,001-point grid candidate in the final selection. The maximum absolute difference between the selected objective and the fine-grid minimum was $2.488\times10^{-6}$ for StateDirect and $3.763\times10^{-7}$ for the feature-free control.

DCC uses three starting pairs and the nested constant-correlation candidate. All 2,304 starts converged across 768 portfolio-month refits, and the forecast correlations were positive definite on the evaluated dates.

\paragraph*{Forecast Timing and Safeguards.}
Feature inputs end on the preceding trading day, and M7 training labels precede the first forecast date of each monthly block. The target normalization uses the training-sample standard deviation, subject to a lower bound of $10^{-8}$, which was never binding in the empirical analysis.

Forecast reproducibility was assessed using truncated return histories at three cutoffs across the four portfolios. Among 260 forecasts from M7, DCC, and the state-dependent and feature-free direct-score rules, the maximum absolute discrepancies in VaR and ES were $2.77\times10^{-11}$ and $3.51\times10^{-11}$, respectively. Repeating the 2010 validation procedure yielded the same M7 specifications for all four portfolios, while parameter re-estimation reproduced all 348 parameter values across twelve sampled portfolio-months. These checks replay selected historical forecasts and parameter fits. A complete historical re-estimation would require refitting at every scheduled date.

Across the three estimation windows, 20 estimation attempts failed. Five scheduled fits in the 504-day window retained the preceding successful parameter estimates after unsuccessful retries. No portfolio-level fallback was required. Correlation estimates were obtained without eigenvalue clipping, diagonal jitter, or additional matrix regularization. All evaluated endpoint and DCC correlation matrices were positive definite, portfolio variance forecasts were strictly positive, and all issued forecasts satisfied $0<q\leq e$.

Independent calculations from the dated forecast panel were used to verify paired FZ0 scores, 420 bootstrap comparisons, and 44 model confidence set entries.

\section{Candidate-Set and Inference Sensitivity}\label{app:mcs}
Table~\ref{tab:adaptive_mcs} reports the main-window confidence sets for M1--M11. Table~\ref{tab:window_summary} summarizes the 180 score contrasts across estimation windows, controls and bootstrap block lengths.
\begin{table}[!htbp]\centering
\caption{Main-window model confidence sets.}\label{tab:adaptive_mcs}
\begingroup\small\singlespacing
\setlength{\tabcolsep}{4pt}
\begin{tabular}{lrrrr}\toprule
Method & P1 & P2 & P3 & P4\\\midrule
M1 HS & No & Yes & Yes & Yes\\
M2 EWMA-N & No & No & No & No\\
M3 GARCH-N & Yes & Yes & Yes & Yes\\
M4 GARCH-$t$ & Yes & Yes & Yes & Yes\\
M5 EQ-$t$ & No & Yes & Yes & Yes\\
M6 EW-$t$ & Yes & Yes & Yes & Yes\\
M7 ML & Yes & Yes & Yes & Yes\\
M8 Equal mixing & Yes & Yes & Yes & Yes\\
M9 Volatility rule & No & Yes & Yes & Yes\\
M10 Score rule & Yes & Yes & Yes & Yes\\
M11 Expanding mean & Yes & No & Yes & Yes\\
\bottomrule\end{tabular}
\par\medskip\footnotesize\textit{Notes:} The 90\% range-statistic MCS uses 1,999 circular resamples with 10-day blocks, 2011--2025, and $W=756$.
\endgroup\end{table}
\begin{table}[!htbp]\centering
\caption{Block-length sensitivity across the twelve portfolio--window cells.}\label{tab:window_summary}
\begingroup\small\singlespacing
\setlength{\tabcolsep}{4pt}
\begin{tabular}{rlrrrrr}\toprule
Block & Control & Cells & Lower & Higher & Includes 0 & Holm lower\\\midrule
10 & M11 Expanding mean & 12 & 1 & 0 & 11 & ---\\
10 & M12 Fixed weight & 12 & 0 & 0 & 12 & ---\\
10 & M5 EQ-$t$ & 12 & 3 & 0 & 9 & ---\\
10 & M6 EW-$t$ & 12 & 0 & 0 & 12 & ---\\
10 & M8 Equal mixing & 12 & 1 & 2 & 9 & 0\\
20 & M11 Expanding mean & 12 & 1 & 1 & 10 & ---\\
20 & M12 Fixed weight & 12 & 0 & 0 & 12 & ---\\
20 & M5 EQ-$t$ & 12 & 3 & 0 & 9 & ---\\
20 & M6 EW-$t$ & 12 & 1 & 0 & 11 & ---\\
20 & M8 Equal mixing & 12 & 1 & 1 & 10 & 0\\
60 & M11 Expanding mean & 12 & 1 & 0 & 11 & ---\\
60 & M12 Fixed weight & 12 & 0 & 0 & 12 & ---\\
60 & M5 EQ-$t$ & 12 & 3 & 0 & 9 & ---\\
60 & M6 EW-$t$ & 12 & 0 & 0 & 12 & ---\\
60 & M8 Equal mixing & 12 & 1 & 0 & 11 & 0\\
\bottomrule\end{tabular}
\par\medskip\footnotesize\textit{Notes:} Lower and higher count pointwise intervals entirely below and above zero, respectively. Holm adjustment covers the twelve M7-minus-M8 contrasts separately at each block length.
\endgroup\end{table}

\section{Calibration and Mechanism Summaries}\label{app:calibration_summary}
Table~\ref{tab:esr_summary} summarizes the 168 ESR tests. Table~\ref{tab:esr_windows} gives M7 $p$-values for the two alternative estimation windows.
\begin{table}[!htbp]\centering
\caption{ESR rejection counts for every evaluated method.}\label{tab:esr_summary}
\begingroup\small\singlespacing
\setlength{\tabcolsep}{4pt}
\begin{tabular}{lrrrr}\toprule
Method & Total tests & Strict rejects & Aux. rejects & M7 Holm rejects\\\midrule
M11 Expanding mean & 24 & 3 & 2 & ---\\
M12 Fixed weight & 24 & 6 & 6 & ---\\
M5 EQ-$t$ & 24 & 10 & 7 & ---\\
M6 EW-$t$ & 24 & 4 & 5 & ---\\
M7 ML & 24 & 2 & 4 & 0\\
M8 Equal mixing & 24 & 2 & 3 & ---\\
M14 Direct FZ0 & 8 & 0 & 1 & ---\\
M15 Constant FZ0 & 8 & 1 & 2 & ---\\
M16 Linear gap & 8 & 0 & 0 & ---\\
\bottomrule\end{tabular}
\par\medskip\footnotesize\textit{Notes:} Counts use pointwise 5\% tests. Only M7 has a twenty-four-test Holm family.
\endgroup\end{table}
\begin{table}[!htbp]\centering
\caption{M7 regression calibration at the additional windows.}\label{tab:esr_windows}
\begingroup\small\singlespacing
\setlength{\tabcolsep}{4pt}
\begin{tabular}{lrrr}\toprule
Portfolio & $W$ & Strict $p$ & Auxiliary $p$\\\midrule
P1 & 504 & 0.065 & 0.055\phantom{0}\\
P1 & 1,008 & 0.068 & 0.073\phantom{0}\\
P2 & 504 & 0.046 & 0.044\phantom{0}\\
P2 & 1,008 & 0.057 & 0.171\phantom{0}\\
P3 & 504 & 0.070 & 0.029\phantom{0}\\
P3 & 1,008 & 0.064 & 0.244\phantom{0}\\
P4 & 504 & 0.022 & 0.022\phantom{0}\\
P4 & 1,008 & 0.063 & 0.0495\\
\bottomrule\end{tabular}
\par\medskip\footnotesize\textit{Notes:} Values are pointwise. None of the twenty-four M7 ESR tests rejects after Holm adjustment.
\endgroup\end{table}

Table~\ref{tab:gain_summary} reports the selected amplification rule's effects relative to M7. Table~\ref{tab:direct_summary} summarizes the 96 direct-objective score contrasts by control and block length.
\begin{table}[!htbp]\centering
\caption{FZ0 differences for validation-selected amplification (M13 versus M7).}\label{tab:gain_summary}
\begingroup\small\singlespacing
\setlength{\tabcolsep}{4pt}
\begin{tabular}{lrrr}\toprule
Portfolio & Amplification $k$ & M13 minus M7 & 20-day 95\% interval\\\midrule
P1 & 0 & 1.631 & \tableinterval{-0.084}{3.363}\\
P2 & 4 & 0.165 & \tableinterval{-2.603}{3.237}\\
P3 & 4 & $-0.470$ & \tableinterval{-2.250}{1.753}\\
P4 & 0 & $-0.327$ & \tableinterval{-0.956}{0.276}\\
\bottomrule\end{tabular}
\par\medskip\footnotesize\textit{Notes:} Differences and interval endpoints are reported in units of $10^{-3}$ FZ0. Amplification is selected in 2010. The displayed intervals use 20-day blocks and also include zero with 10- and 60-day blocks.
\endgroup\end{table}
\begin{table}[!htbp]\centering
\caption{Direct-objective comparison sensitivity.}\label{tab:direct_summary}
\begingroup\small\singlespacing
\setlength{\tabcolsep}{4pt}
\begin{tabular}{rlrrrrr}\toprule
Block & Control & Cells & Lower & Higher & Includes 0 & Holm lower\\\midrule
10 & M11 Expanding mean & 4 & 0 & 0 & 4 & ---\\
10 & M12 Fixed weight & 4 & 0 & 0 & 4 & ---\\
10 & M15 Constant FZ0 & 4 & 0 & 0 & 4 & ---\\
10 & M16 Linear gap & 4 & 0 & 0 & 4 & 0\\
10 & M5 EQ-$t$ & 4 & 0 & 0 & 4 & ---\\
10 & M6 EW-$t$ & 4 & 0 & 0 & 4 & ---\\
10 & M7 ML & 4 & 0 & 0 & 4 & ---\\
10 & M8 Equal mixing & 4 & 0 & 0 & 4 & ---\\
20 & M11 Expanding mean & 4 & 0 & 0 & 4 & ---\\
20 & M12 Fixed weight & 4 & 0 & 0 & 4 & ---\\
20 & M15 Constant FZ0 & 4 & 0 & 0 & 4 & ---\\
20 & M16 Linear gap & 4 & 0 & 0 & 4 & 0\\
20 & M5 EQ-$t$ & 4 & 0 & 0 & 4 & ---\\
20 & M6 EW-$t$ & 4 & 0 & 0 & 4 & ---\\
20 & M7 ML & 4 & 0 & 0 & 4 & ---\\
20 & M8 Equal mixing & 4 & 0 & 0 & 4 & ---\\
60 & M11 Expanding mean & 4 & 0 & 0 & 4 & ---\\
60 & M12 Fixed weight & 4 & 0 & 0 & 4 & ---\\
60 & M15 Constant FZ0 & 4 & 0 & 0 & 4 & ---\\
60 & M16 Linear gap & 4 & 0 & 0 & 4 & 0\\
60 & M5 EQ-$t$ & 4 & 0 & 0 & 4 & ---\\
60 & M6 EW-$t$ & 4 & 0 & 0 & 4 & ---\\
60 & M7 ML & 4 & 0 & 0 & 4 & ---\\
60 & M8 Equal mixing & 4 & 0 & 0 & 4 & ---\\
\bottomrule\end{tabular}
\par\medskip\footnotesize\textit{Notes:} Each row summarizes four M14-minus-control comparisons. Holm adjustment applies to M14 versus M16.
\endgroup\end{table}

\paragraph*{Supplementary Dependence Contrasts.}
Table~\ref{tab:dependence_supplement} gives the supplementary comparisons discussed in Section~\ref{sec:mechanism}.
\begin{table}[!htbp]\centering\footnotesize\singlespacing
\setlength{\tabcolsep}{5pt}
\caption{Supplementary dependence and matched-state FZ0 contrasts.}\label{tab:dependence_supplement}
\begin{tabular}{lrrr@{\hspace{14pt}}rrr}\toprule
& \multicolumn{3}{c}{DCC minus M7}
& \multicolumn{3}{c}{StateDirect minus feature-free control}\\
\cmidrule(lr){2-4}\cmidrule(lr){5-7}
Portfolio & Mean & 95\% interval & Holm $p$ & Mean & 95\% interval & Holm $p$\\\midrule
P1 & 2.489 & \tableinterval[99]{-5.565}{10.550} & 1.000 & 4.963 & \tableinterval[99]{-3.914}{14.232} & 1.000\\
P2 & $-0.670$ & \tableinterval[99]{-5.731}{4.343} & 1.000 & 1.716 & \tableinterval[99]{-6.753}{9.975} & 1.000\\
P3 & $-0.911$ & \tableinterval[99]{-5.034}{3.182} & 1.000 & 0.166 & \tableinterval[99]{-8.918}{8.475} & 1.000\\
P4 & $-1.380$ & \tableinterval[99]{-4.513}{2.241} & 1.000 & $-0.127$ & \tableinterval[99]{-8.188}{7.004} & 1.000\\
\bottomrule\end{tabular}
\par\medskip\footnotesize\textit{Notes:} Differences and pointwise 95\% interval endpoints are reported in units of $10^{-3}$ FZ0. The estimation window is 756 days and the paired bootstrap block length is 20 days. Each Holm adjustment covers the four portfolios for that contrast. Negative differences favor the first rule.
\end{table}

\paragraph*{Exploratory State-Conditional Differences.}
The state analysis uses monthly expanding-sample quartiles of five features to assign low, middle and high states. For each benchmark, this yields 60 overlapping portfolio--feature--state comparisons. The unadjusted pointwise 95\% intervals use 20-day blocks drawn from the full date axis. All differences and interval bounds below are in $10^{-3}$ FZ0 units.

In P1, two intervals favor M7 over equal mixing. The contrasts are $-2.809$ ($[-5.546,\,-0.068]$) in the middle volatility-ratio state and $-7.467$ ($[-14.216,\,-0.817]$) in the high matrix-disagreement state. The interval favors equal mixing in the P3 low volatility-ratio state, with a contrast of $0.844$ ($[0.108,\,1.480]$). For M7 minus validation-selected fixed mixing, the only interval excluding zero is for P2 in the middle matrix-disagreement state, with a contrast of $-10.581$ ($[-18.449,\,-2.640]$). These isolated results from overlapping states are exploratory.

\section{Economic and Simulation Summaries}\label{app:economic_sim_summary}
We assess the economic implications of the forecasts under two hypothetical rules for setting an ES-based buffer.

For model $j$, let $e_{j,t}$ be its ES forecast and $\kappa_j$ its buffer multiplier. The buffer and uncovered loss are
\begin{equation*}
 K_{j,t}=\kappa_j e_{j,t},\qquad G_{j,t}=(L_t-K_{j,t})_+.
\end{equation*}
Here $(x)_+:=\max(x,0)$. The unit-ES rule sets every $\kappa_j=1$. The validation-budget rule sets
\begin{equation*}
 B=\overline e^{\rm HS}_{2010},\qquad \kappa_j=B/\overline e_{j,2010},
\end{equation*}
where $\overline e_{j,2010}$ and $\overline e^{\rm HS}_{2010}$ are averages over the 252 validation dates. The rule equalizes mean validation buffers and holds the multipliers fixed during evaluation.

Let $\overline K_j$ and $\overline G_j$ be time averages over the common evaluation dates. With carrying charge $c_K\geq0$ and uncovered-loss penalty $c_G>0$, define model $j$'s mean economic cost and the cost ratio by
\begin{equation*}
 J_j:=c_K\overline K_j+c_G\overline G_j,\qquad \eta:=c_K/c_G.
\end{equation*}
The normalized mean cost is
\begin{equation}\label{eq:cost}
 J_j/c_G=\eta\overline K_j+\overline G_j,\qquad \eta\in\{0.005,0.01\}.
\end{equation} We use illustrative cost ratios and compare forecasts on common dates with twenty-day circular blocks.

Table~\ref{tab:economic_effects} reports the main-window comparisons with equal mixing using the normalized cost in Equation~\eqref{eq:cost}. Across the twelve portfolio--window cells, the validation-budget rule yields six lower-cost and two higher-cost pointwise intervals at each cost ratio. Under the unit-ES rule, one interval favors M7 and none favors equal mixing at each ratio. With M5 as the comparator, eleven of twelve validation-budget intervals indicate higher cost for M7 at both ratios. Appendix Table~\ref{tab:adaptive_economics} reports all four controls.
\begin{table}[!htbp]\centering
\caption{Main-window M7 cost effects relative to equal mixing.}\label{tab:economic_effects}
\begingroup\small\singlespacing
\setlength{\tabcolsep}{4pt}
\begin{tabular}{llrrr}\toprule
Portfolio & Policy & $\eta$ & Cost difference & 95\% interval\\\midrule
P1 & Unit ES & 0.005 & $-0.500$ & \tableinterval[99]{-3.600}{2.400}\\
P1 & Unit ES & 0.010 & 0.000 & \tableinterval[99]{-3.100}{3.000}\\
P1 & Validation budget & 0.005 & $-8.600$ & \tableinterval[99]{-9.900}{-6.800}\\
P1 & Validation budget & 0.010 & $-16.600$ & \tableinterval[99]{-18.900}{-14.100}\\
P2 & Unit ES & 0.005 & 0.100 & \tableinterval[99]{-2.000}{2.200}\\
P2 & Unit ES & 0.010 & 0.400 & \tableinterval[99]{-1.700}{2.600}\\
P2 & Validation budget & 0.005 & 1.500 & \tableinterval[99]{0.900}{1.900}\\
P2 & Validation budget & 0.010 & 2.900 & \tableinterval[99]{1.900}{3.800}\\
P3 & Unit ES & 0.005 & $-0.300$ & \tableinterval[99]{-1.600}{1.300}\\
P3 & Unit ES & 0.010 & $-0.300$ & \tableinterval[99]{-1.700}{1.300}\\
P3 & Validation budget & 0.005 & 0.200 & \tableinterval[99]{-0.200}{0.600}\\
P3 & Validation budget & 0.010 & 0.400 & \tableinterval[99]{-0.400}{1.200}\\
P4 & Unit ES & 0.005 & 0.200 & \tableinterval[99]{-1.400}{1.800}\\
P4 & Unit ES & 0.010 & 0.200 & \tableinterval[99]{-1.400}{1.800}\\
P4 & Validation budget & 0.005 & 0.500 & \tableinterval[99]{-0.100}{1.100}\\
P4 & Validation budget & 0.010 & 1.100 & \tableinterval[99]{-0.100}{2.200}\\
\bottomrule\end{tabular}
\par\medskip\footnotesize\textit{Notes:} Cost differences and interval endpoints are in units of $10^{-3}$ basis points. Differences are M7 minus M8 normalized cost, $\Delta J/c_G$, relative to exposure. Pointwise intervals use 20-day blocks.
\endgroup\end{table}

The cost comparisons are exploratory. The intervals are unadjusted, and differences in rejection counts across buffer rules do not constitute a test of a policy interaction.

Table~\ref{tab:adaptive_economics} summarizes 192 paired M7 cost contrasts across the buffer rules, cost ratios, controls and portfolio--window cells.
\begin{table}[!htbp]\centering
\caption{Directional evidence for M7 cost differences.}\label{tab:adaptive_economics}
\begingroup\small\singlespacing
\setlength{\tabcolsep}{4pt}
\begin{tabular}{lllrrrr}\toprule
Policy & $\eta$ & Control & Cells & Lower & Higher & Includes 0\\\midrule
Validation budget & 0.005 & M11 Expanding mean & 12 & 5 & 3 & 4\\
Validation budget & 0.005 & M12 Fixed weight & 12 & 3 & 8 & 1\\
Validation budget & 0.005 & M5 EQ-$t$ & 12 & 0 & 11 & 1\\
Validation budget & 0.005 & M8 Equal mixing & 12 & 6 & 2 & 4\\
Validation budget & 0.010 & M11 Expanding mean & 12 & 5 & 3 & 4\\
Validation budget & 0.010 & M12 Fixed weight & 12 & 3 & 8 & 1\\
Validation budget & 0.010 & M5 EQ-$t$ & 12 & 0 & 11 & 1\\
Validation budget & 0.010 & M8 Equal mixing & 12 & 6 & 2 & 4\\
Unit ES & 0.005 & M11 Expanding mean & 12 & 1 & 0 & 11\\
Unit ES & 0.005 & M12 Fixed weight & 12 & 0 & 0 & 12\\
Unit ES & 0.005 & M5 EQ-$t$ & 12 & 2 & 0 & 10\\
Unit ES & 0.005 & M8 Equal mixing & 12 & 1 & 0 & 11\\
Unit ES & 0.010 & M11 Expanding mean & 12 & 1 & 0 & 11\\
Unit ES & 0.010 & M12 Fixed weight & 12 & 0 & 0 & 12\\
Unit ES & 0.010 & M5 EQ-$t$ & 12 & 0 & 0 & 12\\
Unit ES & 0.010 & M8 Equal mixing & 12 & 1 & 0 & 11\\
\bottomrule\end{tabular}
\par\medskip\footnotesize\textit{Notes:} Lower and higher count pointwise 95\% intervals for $J_7-J_{\rm control}$ entirely below and above zero, respectively. Blocks are 20 days.
\endgroup\end{table}

Table~\ref{tab:sim_summary} summarizes full-period results for eight controls and both metrics. Table~\ref{tab:sim_periods} reports the exploratory subperiod comparisons with equal mixing.
\begin{table}[!htbp]\centering
\caption{Full-period simulation evidence across all controls.}\label{tab:sim_summary}
\begingroup\small\singlespacing
\setlength{\tabcolsep}{4pt}
\begin{tabular}{llrrrrr}\toprule
Metric & Control & Cells & Lower & Higher & Includes 0 & Holm lower\\\midrule
ES error & M10 Score rule & 8 & 8 & 0 & 0 & ---\\
ES error & M11 Expanding mean & 8 & 2 & 1 & 5 & ---\\
ES error & M12 Fixed weight & 8 & 3 & 0 & 5 & ---\\
ES error & M5 EQ-$t$ & 8 & 4 & 4 & 0 & ---\\
ES error & M6 EW-$t$ & 8 & 6 & 1 & 1 & ---\\
ES error & Fixed tree specification & 8 & 0 & 0 & 8 & ---\\
ES error & M8 Equal mixing & 8 & 6 & 0 & 2 & ---\\
ES error & M9 Volatility rule & 8 & 4 & 0 & 4 & ---\\
FZ0 & M10 Score rule & 8 & 4 & 0 & 4 & ---\\
FZ0 & M11 Expanding mean & 8 & 0 & 0 & 8 & ---\\
FZ0 & M12 Fixed weight & 8 & 4 & 0 & 4 & ---\\
FZ0 & M5 EQ-$t$ & 8 & 3 & 4 & 1 & ---\\
FZ0 & M6 EW-$t$ & 8 & 6 & 0 & 2 & ---\\
FZ0 & Fixed tree specification & 8 & 0 & 0 & 8 & ---\\
FZ0 & M8 Equal mixing & 8 & 1 & 0 & 7 & 0\\
FZ0 & M9 Volatility rule & 8 & 3 & 0 & 5 & ---\\
\bottomrule\end{tabular}
\par\medskip\footnotesize\textit{Notes:} Each row counts eight scenario--dimension intervals for tuned M7 minus the control. Holm adjustment covers full-period FZ0 contrasts with M8. ES error is oracle-relative ES RRMSE.
\endgroup\end{table}
\begin{table}[!htbp]\centering
\caption{Earlier and later evaluation intervals.}\label{tab:sim_periods}
\begingroup\small\singlespacing
\setlength{\tabcolsep}{4pt}
\begin{tabular}{lrlrrr}\toprule
Mechanism & $d$ & Period & FZ0 diff. & FZ0 interval & ES error interval\\\midrule
Stable N & 2 & Early & $-0.036$ & \tableinterval{-0.391}{0.318} & \tableinterval{-0.835}{-0.262}\\
Stable N & 2 & Late & $-0.082$ & \tableinterval{-0.491}{0.328} & \tableinterval{-0.788}{-0.005}\\
Stable N & 8 & Early & $-0.309$ & \tableinterval{-0.888}{0.269} & \tableinterval{-0.799}{-0.231}\\
Stable N & 8 & Late & $-0.160$ & \tableinterval{-0.674}{0.355} & \tableinterval{-1.256}{-0.348}\\
Stable $t$ & 2 & Early & $-0.401$ & \tableinterval{-0.935}{0.133} & \tableinterval{-0.943}{0.093}\\
Stable $t$ & 2 & Late & $-0.245$ & \tableinterval{-0.726}{0.237} & \tableinterval{-1.137}{-0.296}\\
Stable $t$ & 8 & Early & $-0.328$ & \tableinterval{-0.981}{0.324} & \tableinterval{-0.779}{0.002}\\
Stable $t$ & 8 & Late & $-0.328$ & \tableinterval{-0.905}{0.248} & \tableinterval{-1.049}{-0.319}\\
Switching N & 2 & Down & 0.236 & \tableinterval{-0.173}{0.644} & \tableinterval{-0.437}{0.036}\\
Switching N & 2 & Up & 0.035 & \tableinterval{-0.226}{0.296} & \tableinterval{-0.430}{-0.135}\\
Switching N & 8 & Down & $-0.231$ & \tableinterval{-0.617}{0.154} & \tableinterval{-0.281}{0.259}\\
Switching N & 8 & Up & $-0.210$ & \tableinterval{-0.555}{0.134} & \tableinterval{-0.615}{-0.211}\\
Switching $t$ & 2 & Down & $-0.219$ & \tableinterval{-0.737}{0.300} & \tableinterval{-0.173}{0.446}\\
Switching $t$ & 2 & Up & 0.036 & \tableinterval{-0.227}{0.299} & \tableinterval{-0.124}{0.434}\\
Switching $t$ & 8 & Down & 0.366 & \tableinterval{-0.149}{0.881} & \tableinterval{-0.221}{0.557}\\
Switching $t$ & 8 & Up & 0.093 & \tableinterval{-0.232}{0.418} & \tableinterval{-0.495}{-0.002}\\
\bottomrule\end{tabular}
\par\medskip\footnotesize\textit{Notes:} Differences are tuned M7 minus M8. FZ0 differences and their interval endpoints are reported in units of $10^{-3}$ FZ0. Early or Down covers days 500--749 and Late or Up days 750--999. The stable designs have constant correlation. ES RRMSE differences are changes in dimensionless ratios. They and their interval endpoints are scaled by $10^3$. Intervals are pointwise $t_{49}$ intervals.
\endgroup\end{table}

\section{Component and Validation Timings}\label{app:computation}
Table~\ref{tab:micro_summary} reports median component timings from nine randomized repetitions after warm-up at each of three training sample sizes. The complete evaluation replay and tree validation search each ran once per portfolio. Table~\ref{tab:s_search} reports the validation search and the median of nine fixed-grid timings. The replay excludes warm-up, file handling, and checks. The amortized values describe the timed forecasting components on this machine.
\begin{table}[!htbp]\centering
\caption{Computational medians at three training lengths.}\label{tab:micro_summary}
\begingroup\small\singlespacing
\setlength{\tabcolsep}{4pt}
\begin{tabular}{llrrrrr}\toprule
Portfolio & Forecast origin & $n$ & Fit (ms) & Output (ms) & Feature (ms) & Feature history (s)\\\midrule
P1 & 2011-01-03 & 757 & 808.8 & 4.1 & 0.5 & 1.3\\
P1 & 2018-07-02 & 2,643 & 774.9 & 4.0 & 0.5 & 4.5\\
P1 & 2025-12-01 & 4,507 & 608.4 & 3.8 & 0.4 & 5.8\\
P2 & 2011-01-03 & 757 & 531.4 & 3.0 & 0.4 & 1.1\\
P2 & 2018-07-02 & 2,643 & 594.3 & 3.2 & 0.4 & 3.4\\
P2 & 2025-12-01 & 4,507 & 625.4 & 3.6 & 0.5 & 8.6\\
P3 & 2011-01-03 & 757 & 424.6 & 3.7 & 0.6 & 1.3\\
P3 & 2018-07-02 & 2,643 & 425.3 & 4.5 & 0.6 & 4.6\\
P3 & 2025-12-01 & 4,507 & 286.9 & 2.9 & 0.4 & 6.0\\
P4 & 2011-01-03 & 757 & 279.4 & 2.8 & 0.4 & 1.1\\
P4 & 2018-07-02 & 2,643 & 321.8 & 2.8 & 0.4 & 3.8\\
P4 & 2025-12-01 & 4,507 & 337.1 & 3.0 & 0.4 & 6.2\\
\bottomrule\end{tabular}
\par\medskip\footnotesize\textit{Notes:} $n$ is the training sample size. Output uses prepared inputs. Feature history reconstructs features from historical returns, marginal forecasts and scores.
\endgroup\end{table}
\begin{table}[H]\centering
\caption{Complete validation search costs.}\label{tab:s_search}
\begingroup\small\singlespacing
\setlength{\tabcolsep}{4pt}
\begin{tabular}{lrrrr}\toprule
Portfolio & M7 search (s) & Monthly fits & Complete repeats & Fixed grid (ms)\\\midrule
P1 & 36.5 & 48 & 1 & 6.4\\
P2 & 31.3 & 48 & 1 & 2.8\\
P3 & 26.1 & 48 & 1 & 3.8\\
P4 & 25.5 & 48 & 1 & 2.6\\
\bottomrule\end{tabular}
\par\medskip\footnotesize\textit{Notes:} The fixed grid has 21 weights. Search timings exclude historical feature and endpoint preparation.
\endgroup\end{table}

With prepared inputs, median single-day output times range from 2.8 to 4.5 milliseconds. Four-candidate validation searches take 25.5 to 36.5 seconds. The amortized evaluation cost includes monthly fits and daily risk and feature calculations. Feature-history reconstruction holds historical marginal forecasts fixed.

\section{Positive-Control Derivation and Results}\label{app:positive_control}
\subsection{Population Optimizer and Matched-Gap Construction}
Fix the forecast origin and suppress time subscripts. On any compact positive interval containing the endpoint scales, the realized score along the ray $(q,e)=(c_q\sigma,c_e\sigma)$ is continuous and piecewise differentiable, with a possible kink at $L=c_q\sigma$. Away from that kink its derivative is $1/\sigma-L\mathbf1\{L>c_q\sigma\}/(\alpha c_e\sigma^2)$. The derivative is bounded in absolute value by $C(1+|L|)$ with a finite constant $C$ independent of $L$ and $\sigma$. Continuity across the kink gives the same bound for difference quotients. The conditional loss law has no atom at $c_q\sigma$, and conditional integrability permits dominated convergence, giving Equation~\eqref{eq:scale-score-derivative}. Since both endpoint variances are positive, the derivative of scale with respect to weight is bounded on $[0,1]$. The integrable bound therefore also justifies interchanging conditional expectation and integration of the weight derivative.

For the construction in Section~\ref{sec:surrogate}, write $D(F)=\overline S_F(\sigma_6)-\overline S_F(\sigma_5)$ for the endpoint gap under loss law $F$. Let $F_0=N(0,1)$, $F_a=N(0,0.6^2)$, and $F_b=N(0,1.5^2)$. Fix $\sigma_5=0.75$ and solve $D(F_0)=0$ for $\sigma_6$ on $[1,2]$ with the standard normal forecast constants. Then choose
\begin{equation*}
 p=\frac{D(F_0)-D(F_b)}{D(F_a)-D(F_b)},\qquad F_{\rm mix}=pF_a+(1-p)F_b .
\end{equation*}
Here $0<p<1$. Linearity of the expected score in the law yields $D(F_{\rm mix})=D(F_0)$ with the unrounded parameters. Solving the bracketed scale equation gives weights of approximately 0.247059 and 0.277607 for $F_0$ and $F_{\rm mix}$, respectively. A 101-point weight grid gives 0.25 and 0.28, and direct minimization using numerical quadrature agrees with the root-based weights to within $10^{-6}$.

The same construction with centered variance-standardized $t_5$ laws of standard deviations 1, 0.6 and 1.5 retains the normal forecast constants and the two endpoint scales. Interpolation gives $p\approx0.710923$ and a common gap of approximately $-0.112977$. The optimal weights are approximately 0.315926 and 0.343805, with grid optima 0.32 and 0.34 and the same quadrature agreement. These constructions use known laws for evaluation.

\subsection{Conditional Expected Scores}

Conditional on the forecast information set $\mathcal I_t$, positive-control portfolio loss is a mixture of two centered normal or standardized $t_5$ distributions. Here $\mathcal I_t=\mathcal G_t$ in the matched-occupancy design and $\mathcal I_t=\mathcal F_{t-1}$ in the volatility-driven sensitivity design. The low-correlation component has $\rho_0=-0.6$ and probability $1-p_t$. The high-correlation component has $\rho_1=0.8$ and conditional probability $p_t$. The within-component loss variances are
\begin{equation*}
 v_j=(0.6\times0.012)^2+(0.4\times0.007)^2
       +2(0.6\times0.012)(0.4\times0.007)\rho_j,
 \qquad\rho_j\in\{-0.6,0.8\}.
\end{equation*}
For $j\in\{0,1\}$, define the component scale and standardized threshold by
\begin{equation*}
 s_j=\begin{cases}
       \sqrt{v_j},&\text{normal components},\\
       \sqrt{3v_j/5},&\text{variance-standardized Student-}t_5\text{ components},
     \end{cases}
 \qquad z_j=q/s_j.
\end{equation*}
Here $t_5(\cdot)=f_5(\cdot)$ denotes the PDF of the conventional Student-$t$ distribution with five degrees of freedom; variance standardization is incorporated in $s_j$. The upper truncated first moment is $s_j\phi(z_j)$ under normality or $s_j(5+z_j^2)t_5(z_j)/4$ under Student tails. Subtracting $q$ times the component survival probability gives $A_j(q)=\mathrm E[(L-q)_+\mid j]$. Define the conditional expected score by
\begin{equation*}
 \overline S_t(q,e):=\mathrm E[S_t(q,e)\mid\mathcal I_t].
\end{equation*}
Its exact expression is
\begin{equation}\label{eq:conditional_fz}
 \overline S_t(q,e)=\frac{(1-p_t)A_0(q)+p_tA_1(q)}{\alpha e}
                     +\frac qe+\log(e)-1.
\end{equation}
We use Equation~\eqref{eq:conditional_fz} to evaluate issued forecasts and the oracle benchmarks under the known conditional law.

\subsection{Matched-Occupancy Implementation and Diagnostics}
The matched-occupancy design retains 1,000 initial forecast records and 252 further observations before the 1,000 evaluation dates. The 252-observation window selects the fixed mixture and is part of the history available at the first evaluation fit. Lagged scores require 63 forecast records before training eligibility. The stump learner normalizes the endpoint gap using the training standard deviation. The ridge diagnostic uses six predictors: the signed log variance ratio multiplied by a quadratic basis in the standardized log midpoint variance, and the same three products interacted with the observed driver. The penalty is 0.001 after scaling the columns by training standard deviations. Both fitted gap rules apply the negative sigmoid to their normalized gap predictions. The no-driver comparison sets the three interaction terms to zero and refits on the same histories.

The continuous feasible oracle applies Equation~\eqref{eq:scale-score-derivative} using the true law. A 101-point grid checks the optimizer. The unrestricted oracle solves for the true conditional mixture quantile and truncated-mean ES. Mapping regret compares the exact-gap sigmoid with the feasible oracle. Learner regret uses the fitted stump rule. The difference between those two regrets measures what changes when the estimated gap is replaced by the exact gap at the same historical target scale. It can have either sign.

The feasibility loss compares the constrained oracle with unrestricted VaR and ES. To split this loss, we insert a ray oracle that retains $(q,e)=(c_q\sigma,c_e\sigma)$ for any $\sigma>0$. The feasible-to-ray score difference isolates the endpoint scale interval. The ray-to-unrestricted difference isolates the fixed ratio. At $\beta=3$, the two components are approximately $0.0770$ and $1.04\times10^{-4}$ under normal innovations, and $0.0525$ and $1.90\times10^{-5}$ under Student-$t_5$ innovations. All reported regret intervals are pointwise paired path comparisons.

\begin{table}[!htbp]\centering\footnotesize\singlespacing
\setlength{\tabcolsep}{4pt}
\caption{Matched-occupancy diagnostics.}\label{tab:matched_diagnostics}
\begin{tabular}{lrrrrrrrr}\toprule
Law & $\beta$ & \shortstack{High state\\(\%)} & Gap $R^2$ & \shortstack{Stump\\MSE} & \shortstack{Historical\\MSE} & \shortstack{Variance gap\\(\%)} & \shortstack{Near coincident\\(\%)} & \shortstack{Boundary\\(\%)}\\\midrule
Normal & 0 & 50.25 & $-0.462$ & 0.001506 & 0.001070 & 11.80 & 5.20 & 67.78\\
Normal & 3 & 50.11 & $-0.059$ & 0.007238 & 0.006884 & 11.88 & 5.34 & 99.89\\
$t_5$ & 0 & 50.25 & $-1.308$ & 0.002869 & 0.001385 & 14.51 & 4.17 & 58.28\\
$t_5$ & 3 & 50.11 & $-0.172$ & 0.007701 & 0.006623 & 14.67 & 4.27 & 98.94\\
\bottomrule\end{tabular}
\par\smallskip\begin{minipage}{0.98\linewidth}\scriptsize\textit{Notes:} MSEs are in squared FZ0 units. The historical forecast is the expanding mean. We average $R^2$ across paths after computing it within each path. Variance gap is the absolute endpoint difference divided by midpoint variance. Near-coincident dates have a relative gap below 1\%.\end{minipage}
\end{table}

\begin{table}[!htbp]\centering\footnotesize\singlespacing
\setlength{\tabcolsep}{4pt}
\caption{Paired expected-score losses under matched occupancy.}\label{tab:matched_regret}
\begin{tabular}{lrrrr}\toprule
Law & $\beta$ & Mapping regret & Learner regret & Feasibility loss\\\midrule
Normal & 0 & \shortstack{4.302 (0.112)\\{\tableinterval[99]{4.080}{4.524}}} & \shortstack{4.469 (0.123)\\{\tableinterval[99]{4.226}{4.711}}} & \shortstack{2.565 (0.118)\\{\tableinterval[99]{2.332}{2.798}}}\\
Normal & 3 & \shortstack{21.565 (0.175)\\{\tableinterval[99]{21.218}{21.911}}} & \shortstack{24.272 (0.221)\\{\tableinterval[99]{23.835}{24.710}}} & \shortstack{77.112 (0.224)\\{\tableinterval[99]{76.669}{77.555}}}\\
$t_5$ & 0 & \shortstack{3.481 (0.090)\\{\tableinterval[99]{3.303}{3.659}}} & \shortstack{3.669 (0.102)\\{\tableinterval[99]{3.467}{3.871}}} & \shortstack{0.911 (0.076)\\{\tableinterval[99]{0.759}{1.062}}}\\
$t_5$ & 3 & \shortstack{18.688 (0.175)\\{\tableinterval[99]{18.341}{19.034}}} & \shortstack{20.700 (0.217)\\{\tableinterval[99]{20.271}{21.129}}} & \shortstack{52.505 (0.176)\\{\tableinterval[99]{52.157}{52.852}}}\\
\bottomrule\end{tabular}
\par\smallskip\begin{minipage}{0.98\linewidth}\scriptsize\textit{Notes:} Each entry gives the mean, its Monte Carlo standard error in parentheses, and a pointwise 95\% interval. Units are $10^{-3}$ FZ0.\end{minipage}
\end{table}

\subsection{Volatility-Driven Sensitivity Analysis}
As a sensitivity analysis, we consider a known-margin design in which state probabilities depend on the lagged portfolio volatility ratio. This specification allows unconditional state occupancy to vary with state predictability. The matched-occupancy design in Section~\ref{sec:positive_control} holds unconditional state occupancy fixed. Two assets have constant marginal standard deviations $(0.012,0.007)$, portfolio weights $(0.6,0.4)$ and normal or standardized $t_5$ component innovations. Conditional portfolio loss follows a two-state mixture. For each innovation law, both endpoint forecasts use its corresponding normal or variance-standardized Student-$t_5$ tail constants at $\alpha=0.025$. The feasible forecast family continues to mix variances under common tail constants.

Let $V_{t-1}$ be the ratio of lagged 21-day to 63-day portfolio volatility, already one of M7's twelve features. Let $U_t\overset{\rm i.i.d.}{\sim}\operatorname{Uniform}(0,1)$ be independent of prior history and the return innovations. Define $\operatorname{clip}(x,a,b)=\min\{b,\max\{a,x\}\}$. The state feature, high-correlation-state probability and realized correlation are
\begin{align*}
 Z_{t-1}&=\operatorname{clip}\{8\log V_{t-1},-2,2\},\\
 p_t&=\operatorname{logistic}(\beta Z_{t-1}),\qquad
 \rho_t=-0.6+1.4\,\mathbf1\{U_t<p_t\},
\end{align*}
Here $p_t=\Pr(\rho_t=0.8\mid\mathcal F_{t-1})$. We set $\beta\in\{0,3\}$. With $\beta=0$, the next correlation state is independent of history, although estimation errors can still make the endpoint-score gap predictable. With $\beta=3$, historical volatility predicts the state probability. Feedback from portfolio volatility allows the unconditional state frequencies to differ between the two designs.

Fifty independent paths share innovation draws and uniforms across the four law--coefficient combinations within each path. After 1,000 burn-in and 756 history observations, 1,000 forecast records initialize learning, 252 provide prequential validation and 1,000 provide evaluation. After the 63-day score-history requirement, 937 records are eligible initially and 1,189 at evaluation start. Both estimators use 756-day histories, with an EWMA decay of 0.94. The twelve-feature learner is refitted every 21 days on an expanding sample. The fixed mixture is selected on the 252 validation dates. The learner sees historical returns and endpoint estimates and scores, with the latent states and probabilities reserved for evaluation.

We use paired path-level $t_{49}$ intervals and Holm adjustment across four realized FZ0 comparisons with equal mixing. To assess whether the relative gain increases with predictability, we compare the M7-minus-M8 difference at $\beta=3$ with that at $\beta=0$ on the same path, adjusting separately across the two innovation laws. The coefficient changes both predictability and state occupancy, so this contrast compares two data-generating processes. Expected scores, gap prediction errors and comparisons with the expanding-mean gate and the fixed mixture are secondary.

Both oracle benchmarks use the true conditional loss distribution and are included solely for diagnostic evaluation. The perfect-gap benchmark applies the fitted rule's sigmoid and historical target scale to the exact conditional gap. The feasible-grid benchmark minimizes conditional expected FZ0 over $g=0,0.01,\ldots,1$. Comparing the two assesses the cost of converting the endpoint gap to a mixture weight through the sigmoid. Conditional expected scores are calculated using Equation~\eqref{eq:conditional_fz}.
\begin{table}[!htbp]\centering\footnotesize\singlespacing
\setlength{\tabcolsep}{3pt}
\caption{Known-margin positive control: realized effects and conditional benchmarks.}\label{tab:positive_control}
\begin{tabular}{llrrrrrr}\toprule
Law & $\beta$ & M7 minus M8 & 95\% interval & Holm $p$ & \shortstack{M7\\expected diff.} & \shortstack{Perfect-gap\\expected diff.} & \shortstack{Grid\\expected diff.}\\\midrule
Normal & 0 & $-0.388$ & \tableinterval{-0.639}{-0.137} & 0.009 & $-0.320$ & $-0.424$ & $-4.814$\\
Normal & 3 & $-0.773$ & \tableinterval{-1.116}{-0.430} & $<0.001$ & $-0.672$ & $-3.065$ & $-31.927$\\
$t_5$ & 0 & $-0.502$ & \tableinterval{-1.048}{0.044} & 0.141 & $-0.248$ & $-0.388$ & $-4.064$\\
$t_5$ & 3 & $-0.269$ & \tableinterval{-0.589}{0.051} & 0.141 & $-0.283$ & $-1.974$ & $-24.482$\\
\bottomrule\end{tabular}
\par\smallskip\begin{minipage}{0.98\linewidth}\scriptsize\textit{Notes:} Differences and interval endpoints are reported in units of $10^{-3}$ FZ0. Fifty paths per cell and 1,000 evaluation dates per path. All differences are relative to equal mixing. The four realized FZ0 comparisons form the Holm family. Expected scores use the known conditional distribution.\end{minipage}
\end{table}
Table~\ref{tab:positive_control} reports realized and conditional expected FZ0 results. With predictable states, the adjusted FZ0 comparison favors M7 under normal innovations (difference $-7.73\times10^{-4}$, Holm $p<0.001$), while the Student-$t_5$ comparison is inconclusive (difference $-2.69\times10^{-4}$, Holm $p=0.141$). In the secondary comparisons (Table~\ref{tab:positive_secondary}), the normal design favors M7 over the expanding-mean gate pointwise, and the Student-$t_5$ design favors M7 over validation-selected fixed mixing pointwise.

The normal design with $\beta=0$ also favors M7 over equal mixing after adjustment. The paired interaction tests yield no rejections after adjustment across the two innovation laws (Appendix Table~\ref{tab:positive_interaction}). The evidence for a gain over equal mixing is therefore stronger than the evidence for a larger gain when the next state is predictable.

In the designs with predictable correlation states, the mean path-level conditional-gap $R^2$ is $-0.570$ under normal innovations and $-0.929$ under Student-$t_5$ innovations (Appendix Table~\ref{tab:positive_diagnostics}). Even where mixing gains are evident, the estimated gap has higher prediction error than the historical-mean benchmark on this diagnostic.

The feasible grid attains a lower conditional expected score than the perfect-gap sigmoid under both innovation laws when correlation states are predictable (Table~\ref{tab:positive_oracles}). Exact gap prediction thus leaves room for improvement in the weight rule itself.

\begin{table}[!htbp]\centering\footnotesize\singlespacing
\setlength{\tabcolsep}{3pt}
\caption{Correlation-state probabilities, mixture weights and conditional-gap prediction.}\label{tab:positive_diagnostics}
\begin{tabular}{llrrrrrr}\toprule
Law & $\beta$ & \shortstack{Mean\\$p_t$} & \shortstack{SD\\$p_t$} & \shortstack{High-state\\fraction} & \shortstack{Mean\\weight} & \shortstack{Weight\\SD} & \shortstack{Conditional\\gap $R^2$}\\\midrule
Normal & 0 & 0.500 & 0.000 & 0.497 & 0.492 & 0.023 & $-2.452$\\
Normal & 3 & 0.465 & 0.439 & 0.464 & 0.507 & 0.023 & $-0.570$\\
$t_5$ & 0 & 0.500 & 0.000 & 0.497 & 0.493 & 0.021 & $-3.084$\\
$t_5$ & 3 & 0.450 & 0.445 & 0.450 & 0.502 & 0.021 & $-0.929$\\
\bottomrule\end{tabular}
\par\smallskip\begin{minipage}{0.98\linewidth}\scriptsize\textit{Notes:} Entries average fifty path-level statistics within each design. For each path, conditional-gap $R^2$ is one minus the ratio of the mean squared prediction error of M7 to that of the expanding historical mean benchmark. Both errors use the exact conditional gap as the target over the 1,000 evaluation dates. The reported $R^2$ is the arithmetic mean of the fifty path-level values.\end{minipage}
\end{table}
\begin{table}[!htbp]\centering\footnotesize\singlespacing
\setlength{\tabcolsep}{3pt}
\caption{Secondary M7 comparisons with the expanding-mean gate and validation-selected fixed mixing.}\label{tab:positive_secondary}
\begin{tabular}{lllrrrr}\toprule
Law & $\beta$ & Control & \shortstack{Realized\\difference} & 95\% interval & \shortstack{Expected\\difference} & 95\% interval\\\midrule
Normal & 0 & Expanding mean & $-0.262$ & \tableinterval[99]{-0.497}{-0.027} & $-0.222$ & \tableinterval{-0.299}{-0.145}\\
Normal & 0 & Fixed weight & $-0.808$ & \tableinterval[99]{-4.167}{2.552} & $-1.508$ & \tableinterval{-3.316}{0.300}\\
Normal & 3 & Expanding mean & $-0.446$ & \tableinterval[99]{-0.770}{-0.123} & $-0.418$ & \tableinterval{-0.541}{-0.296}\\
Normal & 3 & Fixed weight & $-0.278$ & \tableinterval[99]{-5.004}{4.447} & $-2.287$ & \tableinterval{-5.101}{0.528}\\
$t_5$ & 0 & Expanding mean & $-0.419$ & \tableinterval[99]{-0.948}{0.111} & $-0.165$ & \tableinterval{-0.241}{-0.089}\\
$t_5$ & 0 & Fixed weight & 0.083 & \tableinterval[99]{-3.177}{3.342} & $-2.414$ & \tableinterval{-4.464}{-0.364}\\
$t_5$ & 3 & Expanding mean & $-0.229$ & \tableinterval[99]{-0.539}{0.080} & $-0.226$ & \tableinterval{-0.324}{-0.129}\\
$t_5$ & 3 & Fixed weight & $-8.593$ & \tableinterval[99]{-13.314}{-3.873} & $-6.841$ & \tableinterval{-9.123}{-4.558}\\
\bottomrule\end{tabular}
\par\smallskip\begin{minipage}{0.98\linewidth}\scriptsize\textit{Notes:} Differences and interval endpoints are reported in units of $10^{-3}$ FZ0. Differences are M7 minus the control. Expected scores integrate the known conditional distribution. Intervals are pointwise paired-path $t_{49}$ intervals.\end{minipage}
\end{table}
\begin{table}[!htbp]\centering\footnotesize\singlespacing
\setlength{\tabcolsep}{3pt}
\caption{Evaluation-only benchmark differences from equal mixing.}\label{tab:positive_oracles}
\begin{tabular}{lllrr}\toprule
Law & $\beta$ & Benchmark & Expected difference & 95\% interval\\\midrule
Normal & 0 & Perfect gap & $-0.424$ & \tableinterval[99]{-0.491}{-0.357}\\
Normal & 0 & Feasible grid & $-4.814$ & \tableinterval[99]{-5.254}{-4.374}\\
Normal & 3 & Perfect gap & $-3.065$ & \tableinterval[99]{-3.228}{-2.902}\\
Normal & 3 & Feasible grid & $-31.927$ & \tableinterval[99]{-32.385}{-31.469}\\
$t_5$ & 0 & Perfect gap & $-0.388$ & \tableinterval[99]{-0.462}{-0.314}\\
$t_5$ & 0 & Feasible grid & $-4.064$ & \tableinterval[99]{-4.407}{-3.721}\\
$t_5$ & 3 & Perfect gap & $-1.974$ & \tableinterval[99]{-2.141}{-1.807}\\
$t_5$ & 3 & Feasible grid & $-24.482$ & \tableinterval[99]{-25.002}{-23.962}\\
\bottomrule\end{tabular}
\par\smallskip\begin{minipage}{0.98\linewidth}\scriptsize\textit{Notes:} Differences and interval endpoints are reported in units of $10^{-3}$ FZ0. Pointwise $t_{49}$ intervals use fifty paired paths. Perfect gap uses the M7 sigmoid and its training target scale. The feasible grid includes 101 weights.\end{minipage}
\end{table}
\begin{table}[!htbp]\centering\footnotesize\singlespacing
\setlength{\tabcolsep}{3pt}
\caption{Paired change in the realized M7-minus-M8 effect with state predictability.}\label{tab:positive_interaction}
\begin{tabular}{lrrr}\toprule
Law & Change & 95\% interval & Holm $p$\\\midrule
Normal & $-0.385$ & \tableinterval{-0.764}{-0.007} & 0.093\\
$t_5$ & 0.233 & \tableinterval{-0.365}{0.832} & 0.437\\
\bottomrule\end{tabular}
\par\smallskip\begin{minipage}{0.98\linewidth}\scriptsize\textit{Notes:} Differences and interval endpoints are reported in units of $10^{-3}$ FZ0. Change is the paired M7-minus-M8 difference at $\beta=3$ less the difference at $\beta=0$. Negative values favor a larger relative M7 gain with state predictability. Holm adjustment covers the two innovation laws.\end{minipage}
\end{table}

\FloatBarrier
\paragraph*{Learner Scores at Each Information Strength.}
Table~\ref{tab:matched_learner_half} reports the cell-level score differences for the matched-occupancy design in Section~\ref{sec:positive_control}.
\begin{table}[!htbp]\centering\footnotesize\singlespacing
\setlength{\tabcolsep}{5pt}
\caption{Matched-occupancy learner score differences from equal mixing.}\label{tab:matched_learner_half}
\begin{tabular}{lrlrr@{\hspace{14pt}}rr}\toprule
& & & \multicolumn{2}{c}{Boosted stump} & \multicolumn{2}{c}{Ridge with driver interactions}\\
\cmidrule(lr){4-5}\cmidrule(lr){6-7}
Law & $\beta$ & FZ0 & Mean & 95\% interval & Mean & 95\% interval\\\midrule
Normal & 0 & Realized & $-0.288$ & \tableinterval{-0.410}{-0.165} & $-0.556$ & \tableinterval{-0.844}{-0.268}\\
Normal & 0 & Expected & $-0.265$ & \tableinterval{-0.313}{-0.217} & $-0.366$ & \tableinterval{-0.451}{-0.281}\\
Normal & 3 & Realized & $-0.205$ & \tableinterval{-0.328}{-0.082} & $-2.369$ & \tableinterval{-2.744}{-1.995}\\
Normal & 3 & Expected & $-0.223$ & \tableinterval{-0.264}{-0.182} & $-2.648$ & \tableinterval{-2.821}{-2.474}\\
$t_5$ & 0 & Realized & $-0.345$ & \tableinterval{-0.549}{-0.141} & $-0.300$ & \tableinterval{-0.736}{0.137}\\
$t_5$ & 0 & Expected & $-0.201$ & \tableinterval{-0.256}{-0.147} & $-0.149$ & \tableinterval{-0.262}{-0.037}\\
$t_5$ & 3 & Realized & $-0.070$ & \tableinterval{-0.172}{0.031} & $-1.195$ & \tableinterval{-1.579}{-0.810}\\
$t_5$ & 3 & Expected & $-0.154$ & \tableinterval{-0.202}{-0.106} & $-1.593$ & \tableinterval{-1.777}{-1.408}\\
\bottomrule\end{tabular}
\par\medskip\footnotesize\textit{Notes:} Differences are learner minus equal mixing, in $10^{-3}$ FZ0 units. The pointwise intervals use 128 paired paths. Expected scores are evaluated under the true conditional loss law.
\end{table}

\end{document}